\documentclass[aps,prd,twocolumn,floatfix,nofootinbib,superscriptaddress,showpacs,amssymb]{revtex4}

\usepackage{bm}
\usepackage{graphicx}

\newcommand{\der}[2]{\frac{\partial #1}{\partial #2}}
\newcommand{\dert}[2]{{\partial #1 / \partial #2}}
\newcommand{\dder}[2]{\frac{\partial^2 #1}{\partial #2 ^2}}
\newcommand{\w}[1]{\bm{#1}}
\newcommand{\be}{\begin{equation}}
\newcommand{\ee}{\end{equation}}
\newcommand{\bea}{\begin{eqnarray}}
\newcommand{\eea}{\end{eqnarray}}
\let\gm\gamma
\let\ph\varphi
\let\th\theta
\newcommand{\tgm}{\tilde\gm}
\newcommand{\cD}{{\cal D}}
\newcommand{\tna}{\tilde D}
\newcommand{\taa}{\tilde A}

\begin{document}


\title{A constrained scheme for Einstein equations based on
Dirac gauge and spherical coordinates}



\author{Silvano Bonazzola}
\email[]{Silvano.Bonazzola@obspm.fr}

\author{Eric Gourgoulhon}
\email[]{Eric.Gourgoulhon@obspm.fr}
\affiliation{Laboratoire de l'Univers et de ses Th\'eories,
UMR 8102 du C.N.R.S., Observatoire de Paris, F-92195 Meudon Cedex,  France}

\author{Philippe Grandcl\'ement}
\email[]{grandcle@phys.univ-tours.fr}
\affiliation{Laboratoire de Math\'ematiques et de Physique Th\'eorique,
UMR 6083 du C.N.R.S.,
Universit\'e de Tours, Parc de Grandmont, F-37200 Tours, France}
\affiliation{Laboratoire de l'Univers et de ses Th\'eories,
UMR 8102 du C.N.R.S., Observatoire de Paris, F-92195 Meudon Cedex, France}

\author{J\'er\^ome Novak}
\email[]{Jerome.Novak@obspm.fr}
\affiliation{Laboratoire de l'Univers et de ses Th\'eories,
UMR 8102 du C.N.R.S., Observatoire de Paris, F-92195 Meudon Cedex, France}


\date{6 September 2004}

\begin{abstract}
We propose a new formulation for 3+1 numerical relativity, based on a
constrained scheme and a generalization of Dirac gauge to spherical
coordinates. This is made possible thanks to the introduction 
of a flat 3-metric on the spatial
hypersurfaces $t={\rm const}$, which corresponds to the asymptotic
structure of the physical 3-metric induced by the spacetime metric. 
Thanks to the joint use of Dirac gauge, maximal slicing and 
spherical components of tensor fields, the ten Einstein equations are 
reduced to a system of five quasi-linear elliptic equations
(including the Hamiltonian and momentum constraints) 
coupled to two quasi-linear scalar wave equations. The
remaining three degrees of freedom are fixed by
the Dirac gauge. Indeed this gauge allows a direct computation of the 
spherical components of the conformal metric from the two scalar potentials
which obey the wave equations.  
We present some numerical evolution of 3-D gravitational wave 
spacetimes which 
demonstrates the stability of the proposed scheme.  
\end{abstract}

\pacs{04.20.Ex,04.20.Cv,04.25.Dm,04.30.Db}

\maketitle


\section{Introduction and motivations}

Motivated by the construction of the detectors LIGO, GEO600, TAMA and
VIRGO, as well as by the space project LISA,
numerical studies of gravitational wave sources are numerous
(see \cite{BaumgS03,Lehne01} for recent reviews). The majority of them are
performed within the framework of the so-called {\em 3+1 formalism} of general
relativity, also called {\em Cauchy formulation}, in which the spacetime 
is foliated by a family of
spacelike hypersurfaces.
We propose here a new strategy within this formalism,
based on a constrained scheme and spherical coordinates,
which is motivated as follows.

\subsection{Motivations for a constrained scheme} \label{s:mot_constr}

In the 3+1 formalism, the Einstein equations are decomposed in a
set of four {\em constraint equations} and a set of six
{\em dynamical equations} \cite{York79,BaumgS03}. 
The constraint equations give rise
to elliptic (or sometime parabolic) partial differential equations (PDE), 
whereas the PDE type of the dynamical equations depends on the choice of
the coordinate system. Various strategies can then be contemplated:
(i) {\em free evolution scheme:}  solving the constraint
equations only to get the initial data and performing the time
evolution via the dynamical equations, without
enforcing the constraints; (ii)
{\em partially constrained scheme:} using some of the constraints
to compute some of the metric components during the evolution and
(iii) {\em fully constrained scheme:} solving the four constraint equations
at each time step. 

In the eighties, partially constrained schemes, 
with only the Hamiltonian constraint enforced,
have been widely used in 2-D (axisymmetric) computations 
(e.g. Bardeen and Piran \cite{BardeP83}, 
Stark and Piran \cite{StarkP85}, Evans \cite{Evans86}).
Still in the 2-D axisymmetric case, 
fully constrained schemes have been used 
by Evans \cite{Evans89}
and Shapiro and Teukolsky \cite{ShapiT92} 
for non-rotating spacetimes, and by Abrahams, Cook, 
Shapiro and Teukolsky \cite{AbrahCST94} for rotating ones. 
We also notice that the recent (2+1)+1 axisymmetric code of Choptuik et al. 
\cite{ChoptHLP03} is based on a constrained scheme too. 

Regarding the 3-D case, almost all numerical studies to date
are based on free evolution schemes\footnote{an 
exception is the recent work \cite{AnderM03}, where some constrained evolution
of a single isolated black hole is presented.}. 
It turned out that the free evolution scheme directly 
applied to the standard 3+1 equations (sometimes called {\em ADM formulation})
failed due to the development of constraint-violating modes.
An impressive amounts of works have then been devoted these last
years to finding stable evolution schemes  
(see \cite{ShinkY03} for an extensive review and \cite{LindbSKPST04}
for a very recent work in this area). Among them, a large number of 
authors have tried to introduce coordinates and auxiliary variables
so that the dynamical equations become a first-order symmetric hyperbolic 
system. However these approaches have revealed very limited success in 
practice. Another approach has become very popular in the last few years:
the so-called {\em BSSN formulation}, originally devised by
Shibata and Nakamura \cite{ShibaN95} and re-introduced by 
Baumgarte and Shapiro \cite{BaumgS99}. It has shown a much improved 
stability with respect to the standard ADM formulation. Indeed the most
successful computations in numerical relativity to date are based on that
formulation (e.g. \cite{ShibaU00,ShibaU01}).

All the approaches mentioned above favor first-order hyperbolic equations 
with respect to elliptic equations. In particular, they employ a 
free-evolution scheme, avoiding to solve the (elliptic) constraint equations. 
The main reason is neither mathematical nor physical, but rather 
a technical one: for most numerical techniques, solving elliptic equations is 
CPU time expensive. In this article, 
we present an approach which is based on the opposite strategy, 
namely to use as much as possible elliptic equations and as few 
hyperbolic equations as possible. More precisely we propose to use a fully 
constrained-evolution scheme and to solve the minimum number 
of hyperbolic equations: 
the two wave equations corresponding
to the two degrees of freedom of the gravitational field.
The main advantages of this procedure are that (i) elliptic equations 
are much more stable than hyperbolic ones, in particular their 
mathematical well-posedness is usually established, (ii) 
the constraint-violating 
modes that plague the free-evolution schemes do not exist by construction 
in a fully constrained evolution, (iii) the equations describing stationary
spacetimes are usually elliptic and are naturally recovered when taking
the steady-state limit of the proposed scheme. Besides, let us point that 
some very efficient (i.e. requiring a modest CPU time) numerical techniques 
(based on spectral methods)
are now available to solve elliptic equations \cite{BonazM90,GrandBGM01}.
Very recently some scheme has been proposed in which the
constraints, re-written as time evolution equations, are satisfied
up to the time discretization error \cite{GentlGKM03}. On the contrary, 
our scheme guarantees that the constraints are fulfilled within the
precision of the {\em space} discretization error (which can have a much
better accuracy, thanks to the use of spectral methods). 

To achieve this aim, we use maximal slicing, as long as a generalization
of Dirac gauge to curvilinear coordinates. This gauge fixes the 
spatial coordinates $(x^i)$ in each hypersurface $t={\rm const}$.
It has been introduced by Dirac in 1959 \cite{Dirac59} as a way
to fix the coordinates in the Hamiltonian formulation of general
relativity, prior to its quantization (see \cite{Deser03} for a
discussion). Dirac gauge has been discussed in the context of numerical
relativity first by Smarr and York, in their search for a radiation 
gauge in general relativity \cite{SmarrY78a}. But they disregarded it
as being not covariant under coordinate transformation $(x^i)\mapsto (x^{i'})$
in the hypersurface $t={\rm const}$. 
They preferred the {\em minimal distortion gauge}, 
which is fully covariant and allows for an arbitrary choice of the coordinates
$(x^i)$ in the initial hypersurface. Here we show that if one introduces
a flat 3-metric on each spatial hypersurface, in addition 
to the physical 3-metric induced by the spacetime metric, the
Dirac gauge can be made covariant. This enables the use of curvilinear
coordinates, whereas Dirac original formulation was only for Cartesian 
coordinates.
However, contrary to the minimal distortion gauge,
this generalized Dirac gauge still determines fully the coordinates
in the initial slice (up to some inner boundary conditions if the 
slice contains some holes).  

\subsection{Motivations for spherical coordinates} \label{s:motiv_sphere}

Since the astrophysical objects we want to model (neutron stars and 
black holes) have spherical topology, it is natural to use
spherical coordinates $(x^i)=(r,\th,\ph)$ to describe them. 
In particular, spherical coordinates and spherical components
of tensor fields 
enable one to treat properly the boundary conditions (i) at the surface
of fluid stars, as well as at some black hole (apparent) horizon,
and (ii) at spatial infinity or at the edge of the computational domain. 
For a binary system, two systems of spherical 
coordinates (each centered on one of the objects) have proved to
be successful in the treatment of binary neutron stars \cite{GourgGTMB01}
and binary black holes \cite{GrandGB02}.

\subsubsection{Outer boundary conditions}
 
For elliptic equations, spherical coordinates allow a natural
$1/r$ compactification which permits to impose boundary conditions 
at spatial infinity \cite{BonazGM98,GrandBGM01}. 
In this way, the imposed boundary conditions are exact. 

For wave equations from a central source, a spherical boundary of 
the numerical domain 
of integration allows to set non-reflecting boundary conditions
\cite{NovakB04}. Moreover the use of spherical components of the
metric tensor allows, in the Dirac gauge, an easy extraction of the
wave components. This results from the asymptotic transverse and traceless
(TT) behavior of Dirac gauge and the fact that a TT tensor wave
propagating in the radial direction is well described with spherical
components.

\subsubsection{Black hole excision}

Spherical coordinates clearly facilitate black hole excision.
Moreover for stationary problems, one has usually to set
the lapse to zero on some sphere $r={\rm const}$, in order
to preserve the time-independent behavior of slicing of stationary
spacetimes \cite{GourgGB02,HannaECB03}.
As we discuss in Appendix~\ref{s:excision}, using spherical components of the metric tensor
and shift vector is crucial is setting boundary condition on an
excised 2-sphere with vanishing lapse function. In fact, because of
the degeneracy of the operator acting on the above quantities when
the lapse is zero, one can impose boundary conditions on certain
components, and not on the others. In Cartesian components (i.e. 
linear combinations of spherical components), the imposition of 
boundary conditions could not be done simply. 

\subsubsection{Fulfilling the Dirac gauge}

We will show that, when using spherical coordinates, the Dirac gauge
condition can be imposed easily on spherical {\em components} of the 
metric tensor. Indeed, we propose to use the Dirac gauge to compute 
directly some metric components from the other ones. This seems difficult
with Cartesian components (even with spherical coordinates).

\subsubsection{Spherical coordinates and numerical techniques}

Despite the above strong advantages and although they
have been widely used for 2-D (axisymmetric) computations 
\cite{BardeP83,StarkP85,Evans86,Evans89,ShapiT92,AbrahCST94,NakamOK87,BrandS95a,BrandS95b}, 
spherical coordinates
are not well spread in 3-D numerical relativity.
A few exceptions are the time evolution of pure gravitational wave
spacetimes by Nakamura et al. \cite{NakamOK87} \footnote{Note that while
Nakamura et al. \cite{NakamOK87} used spherical coordinates, 
they considered Cartesian components of the tensor fields.} and 
the attempts of computing 3-D stellar core
collapse by Stark \cite{Stark89}. 
This situation is mostly due to the massive
usage of finite difference methods, which have difficulties to 
treat the coordinate singularities on the axis $\th=0$ and $\th=\pi$,
and at the origin $r=0$. On the contrary,
spectral methods employed mostly in our group \cite{BonazGM99b,GrandBGM01} 
and Cornell group \cite{PfeifKST03}, deal without any difficulty with the singularities
inherent to spherical coordinates.
Let us note that in other fields of numerical simulation, 
like stellar hydrodynamics,
spherical coordinates are well spread, for instance in the treatment
of supernovae \cite{DimmeFM02,DimmeNFIM04}.

\subsection{Plan of the paper}

We start the present study by introducing in Sec.~\ref{s:cov3p1} 
a conformal decomposition of 
the 3+1 Einstein equations which is fully covariant with respect to 
a background flat metric. This differs slightly from previous 
conformal decompositions (e.g. \cite{ShibaN95,BaumgS99}) by the
fact that our conformal metric is a genuine tensor field, and not
a tensor density. Then in Sec.~\ref{s:3p1flat} 
we re-write the conformal 3+1 Einstein equations in terms of the
covariant derivative with respect to the flat background metric. 
This enables us to introduce the (generalized) Dirac gauge
in Sec.~\ref{s:Dirac}
and to simplify accordingly the equations. 
We introduce as the basic object of our formulation the
difference $\w{h}$ between the inverse conformal metric and the
inverse flat metric. At the end of Sect.~\ref{s:Dirac},
we present an explicit wave equation for $\w{h}$. 
In Sec.~\ref{s:spher}, we introduce spherical coordinates and 
explicit the equations in terms of tensor components with respect 
to an orthonormal spherical 
frame. We show how the Dirac gauge can then be used to deduce
some metric components from the others in a quasi-algebraic
way. The resolution of the dynamical 3+1 equations is then
reduced to the resolution of two (scalar) wave equations. 
A numerical application is presented in Sec.~\ref{s:num}, 
where it is shown that the proposed scheme can evolve stably
pure gravitational wave spacetimes. 
Finally Sec.~\ref{s:concl} gives the concluding remarks. 
This article is intended to be followed by another study
which focuses on the treatment of
boundary conditions at black hole horizon(s). 
Here we present only in Appendix~\ref{s:excision} a preliminary 
discussion about the type and the number of inner boundary conditions
for black hole spacetimes. 

\section{Covariant 3+1 conformal decomposition} \label{s:cov3p1}

\subsection{3+1 formalism} \label{s:3p1form}

We refer the reader to \cite{BaumgS03} and \cite{York79} for
an introduction to the 3+1 formalism of general relativity. 
Here we simply summarize a few key equations, in order mainly to
fix the notations\footnote{We use geometrized units for which
$G=1$ and $c=1$; Greek indices run in $\{0,1,2,3\}$, whereas 
Latin indices run in $\{1,2,3\}$ only.}. 
The spacetime (or at least the part of it under study)
is foliated by a family of spacelike hypersurfaces $\Sigma_t$,
labeled by the time coordinate $t$.
We denote by $\w{n}$ the future directed unit normal to $\Sigma_t$.
By definition
$\w{n}$, considered as a 1-form, is parallel to the gradient of $t$:
\be \label{e:def_n}
	\w{n} = - N {\bf d} t . 
\ee
The proportionality factor $N$ is called the {\em lapse function}.
It ensures that $\w{n}$ satisfies to the normalization relation
$n_\mu n^\mu = - 1$.

The metric $\w{\gm}$ induced by the spacetime metric $\w{g}$ 
onto each hypersurface $\Sigma_t$
is given by the orthogonal projector onto $\Sigma_t$:
\be \label{e:def_3-metric}
	\w{\gm} := \w{g} + \w{n} \otimes \w{n} . 
\ee 
Since $\Sigma_t$ is assumed to be spacelike, $\w{\gm}$ is
a positive definite Riemannian metric. In the following, we
call it the {\em 3-metric} and denote by 
$\w{D}$ the covariant derivative
associated with it. The second fundamental tensor characterizing the 
hypersurface
$\Sigma_t$ is its {\em extrinsic curvature} $\w{K}$,
given by the Lie derivative of
$\w{\gm}$ along the normal vector $\w{n}$:
\be \label{e:def_courb_extrins}
	\w{K} := - {1\over 2}\pounds_{\w{n}}\w{\gm} .
\ee

One introduces on each hypersurface $\Sigma_t$ a coordinate
system $(x^i) = (x^1,x^2,x^3)$ 
which varies smoothly between neighboring
hypersurfaces, so that $(x^\alpha)=(t,x^1,x^2,x^3)$ constitutes a 
well-behaved coordinate system of the whole spacetime\footnote{later on 
we will specify the coordinates $(x^i)$ to be of 
spherical type, with $x^1=r$, $x^2=\th$ and $x^3=\ph$, but
at the present stage we keep $(x^i)$ fully general.}.  
We denote by 
$\left( \dert{}{x^\alpha} \right) = \left( \dert{}{t}, \dert{}{x^i} \right) 
    = \left( \dert{}{t}, \dert{}{x^1}, \dert{}{x^2}, \dert{}{x^3} \right)$
the natural vector basis associated with this coordinate system. 
The 3+1 decomposition of the basis vector $\partial /\partial t$ defines
the {\em shift vector} $\w{\beta}$ of the spatial coordinates
($x^i$):
\be \label{e:dsdt_ortho}
	\der{}{t} 
		= N \w{n} + \w{\beta}
			\qquad \mbox{with} \qquad \w{n}\cdot\w{\beta} = 0 .
\ee
The metric components $g_{\alpha\beta}$ with respect to the
coordinate system $(x^\alpha)$ are expressed in terms of the lapse function
$N$, the shift vector components $\beta^i$ and the 3-metric components
$\gamma_{ij}$ according to 
\be \label{e:line_g}
	g_{\mu\nu} \, dx^\mu\, dx^\nu
	= - N^2 dt^2 + \gamma_{ij} (dx^i + \beta^i dt)
		(dx^j + \beta^j dt) . 
\ee

In the 3+1 formalism, the matter energy-momentum tensor $\w{T}$ 
is decomposed as 
\be
	\w{T} = E\, \w{n}\otimes \w{n} + \w{n} \otimes \w{J}
		+ \w{J}\otimes \w{n} + \w{S} , 
\ee
where the energy density $E$, the momentum density $\w{J}$ and
the strain tensor $\w{S}$, all of them as measured by the observer
of 4-velocity $\w{n}$, are given by the following projections:
$E := T_{\mu\nu} n^\mu n^\nu$, 
$J_\alpha  :=  - \gm_\alpha^{\ \,\mu} T_{\mu\nu} n^\nu$,
$S_{\alpha\beta}  :=  \gm_\alpha^{\ \,\mu} \gm_\beta^{\ \,\nu} T_{\mu\nu}$.
By means of the Gauss and Codazzi relations, the Einstein field equation
is equivalent to the following system of equations
(see e.g. Eqs.~(23), (24) and (39) of York \cite{York79}):
\be \label{e:ham_constr0}
	R + K^2 - K_{ij} K^{ij} = 16\pi E ,
\ee
\be \label{e:mom_constr0}
	D_j K_i^{\ \, j} - D_i K = 8\pi J_i,
\ee
\bea 
	\der{}{t}K_{ij} & - & \pounds_{\w{\beta}} K_{ij} =  - D_i D_j N 
		+ N \big[ R_{ij} - 2 K_{ik} K^k_{\ j} \nonumber \\*
        & & + K K_{ij}
			+ 4\pi \left( (S-E)\gm_{ij} - 2 S_{ij} \right)
			\big] . \label{e:evol_K0}
\eea 
Equation~(\ref{e:ham_constr0}) is called
the {\em Hamiltonian constraint}, Eq.~(\ref{e:mom_constr0})
the {\em momentum constraint} and Eqs.~(\ref{e:evol_K0}) the {\em dynamical
equations}. In these equations $K$ denotes the trace of the extrinsic
curvature: $K:=K^i_{\ \, i}$, $S:=S^i_{\ \, i}$, 
$R_{ij}$ the Ricci tensor
associated with the 3-metric $\w{\gamma}$ and 
$R:=R^i_{\ \, i}$ the corresponding scalar curvature.
These equations must
be supplemented by the kinematical relation (\ref{e:def_courb_extrins})
between $\w{K}$ and $\w{\gm}$:
\be \label{e:kin0}
	\der{}{t}\gm_{ij} - \pounds_{\w{\beta}} \gm_{ij} 
	= - 2N K_{ij}  .	
\ee

\subsection{Conformal metric}

York \cite{York72} has shown that the  dynamical degrees of freedom
of the gravitational field are carried by the conformal ``metric''
$\w{\hat\gm}$ defined by 
\be
    \hat \gm_{ij} := \gm^{-1/3} \, \gm_{ij},    \label{e:def_hatg}
\ee
where
\be \label{e:def_detg}
	\gm := \det\gm_{ij}.
\ee
The quantity defined by Eq.~(\ref{e:def_hatg}) is a tensor density of
weight $-2/3$, which has unit determinant and which is invariant in
any conformal transformation of $\gm_{ij}$. It can be seen as
representing the equivalence class of conformally related metrics
to which the 3-metric $\w{\gm}$ belongs.
The conformal ``metric'' (\ref{e:def_hatg}) has been used notably 
in the BSSN formulation \cite{ShibaN95,BaumgS99}, along with an 
``associated'' covariant derivative $\w{\hat D}$. However, since 
$\w{\hat\gm}$ is a tensor density and not a tensor field, there
is not a unique covariant derivative associated with it. In particular
one has $\w{D}\w{\hat\gm}=0$, so that the covariant derivative 
$\w{D}$ introduced in Sec.~\ref{s:3p1form}
is ``associated'' with $\w{\hat\gm}$, in addition to $\w{\hat D}$.
As a consequence, some of the formulas presented in Refs.~\cite{ShibaN95}, 
\cite{BaumgS99} or \cite{AlcubBDKPST03} have a meaning only for 
Cartesian coordinates. 

To clarify the meaning of $\w{\hat D}$ and to allow for the use of 
spherical coordinates, we introduce an extra structure on the
hypersurfaces $\Sigma_t$, namely a metric $\w{f}$ 
with the following properties: (i) $\w{f}$ has a vanishing 
Riemann tensor (flat metric), (ii) $\w{f}$ does not vary from one 
hypersurface to the next one along the
spatial coordinates lines:
\be \label{e:f_notime}
	\der{}{t} f_{ij} = 0 ,
\ee
and (iii) the asymptotic structure of the physical metric $\w{\gm}$ 
is given by $\w{f}$:
\be
	\gamma_{ij} \sim f_{ij} \qquad \mbox{at spatial infinity}.
\ee
This last relation expresses the asymptotic flatness of the
hypersurfaces $\Sigma_t$, which we assume in this article.

The inverse metric is denoted by $f^{ij}$~\footnote{Note that, in general
one has $f^{ij} \not= \gm^{ik} \gm^{jl}\, f_{kl}$.}:
$ f^{ik} f_{kj} = \delta^i_{\ \, j}$.
We denote by $\w{\cD}$ the unique covariant derivative associated
with $\w{f}$: $\cD_k f_{ij} = 0$
and define
\be \label{e:def_upcD}
	\cD^i := f^{ij} \cD_j . 
\ee
Thanks to the flat metric $\w{f}$, we can properly define the
{\em conformal metric} $\w{\tgm}$ as
\be \label{e:def_tgm2}
	\tgm_{ij} := \Psi^{-4}\, \gm_{ij}
	\qquad \mbox{or} \qquad \gm_{ij} =: \Psi^4 \, \tgm_{ij} ,
\ee
where the conformal factor $\Psi$ is defined by
\be
	\Psi := \left( {\gm\over f} \right)^{1/ 12} ,
\ee
$\gm$ and $f$ being respectively the determinant of $\w{\gm}$ 
[cf. Eq.~(\ref{e:def_detg})] and the determinant of
$\w{f}$ with respect to the coordinates $(x^i)$:
\be \label{e:def_detf}
    f := \det f_{ij}.	
\ee
Being expressible as the quotient of two determinants, $\Psi$ is a
scalar field on $\Sigma_t$. 
Indeed a change of coordinates $(x^i) \mapsto (x^{i'})$ induces the following
changes in the determinants: $\gm' = (\det J)^2 \gm$ and
$f' =  (\det J)^2 f$, where $J$ denotes the Jacobian matrix
$J^i_{\ \, i'} := \dert{x^i}{x^{i'}}$. It is then obvious
that $\gm'/f' = \gm /f$, which shows the covariance of $\gm/f$.
Since $\Psi$ is a scalar field, $\w{\tgm}$ defined by Eq.~(\ref{e:def_tgm2})
is a tensor field
on $\Sigma_t$ and not a tensor density as the quantity defined by
Eq.~(\ref{e:def_hatg}) and considered in the BSSN formulation
\cite{ShibaN95,BaumgS99,BaumgS03}. Moreover, 
$\Psi$ being always strictly positive (for $\gm$ and $f$ are 
strictly positive), $\w{\tgm}$ is a Riemannian metric on $\Sigma_t$.
Actually it is the member of the conformal equivalence class of $\w{\gm}$
which has the same determinant as the flat metric $\w{f}$:
\be \label{e:dettgm_f}
	\det \tgm_{ij} = f .
\ee
In this respect, our approach agrees with the point of view of York
in Ref.~\cite{York99}, who prefers to introduce a specific member of 
the conformal equivalence class of $\w{\gm}$ instead of manipulating
tensor densities such as (\ref{e:def_hatg}). In our case, we 
use the extra-structure $\w{f}$ to pick out
the representative member of the conformal equivalence class by the
requirement (\ref{e:dettgm_f}).

We define the {\em inverse conformal metric} $\tgm^{ij}$ by
the requirement
\be
	\tgm_{ik} \, \tgm^{kj} = \delta_i^{\ \, j} ,
\ee
which is equivalent to
\be
	\tgm^{ij}  = \Psi^4 \, \gm^{ij} \qquad \mbox{or} \qquad 
	\gm^{ij}  = \Psi^{-4} \, \tgm^{ij}.
\ee
Since $\w{\tgm}$ is a well defined metric on $\Sigma_t$, there is a
unique covariant derivative associated with it, which we denote 
by $\w{\tna}$: $\tna_k \tgm_{ij} = 0$.
The covariant derivatives $\w{\tna}\w{T}$ and 
$\w{\cD}\w{T}$ of any tensor field $\w{T}$ of type
$\left({p\atop q}\right)$ on $\Sigma_t$ are related by the formula
\bea 
	\tna_k T^{i_1\ldots i_p}_{\quad \quad j_1\ldots j_q}
	& = & \cD_k T^{i_1\ldots i_p}_{\quad \quad j_1\ldots j_q}
	+ \sum_{r=1}^p \Delta^{i_r}_{\ \, lk} \, 
		T^{i_1\ldots l \ldots i_p}_{\quad\quad \quad j_1\ldots j_q}
        \nonumber \\*
	& & - \sum_{r=1}^q \Delta^l_{\ \, j_r k} \, 
		T^{i_1\ldots i_p}_{\quad \quad j_1\ldots l \ldots j_q} ,
            \label{e:def_nablatilde}
\eea  
where $\w{\Delta}$ denotes the following type $\left({1\atop 2}\right)$ tensor field:
\be \label{e:def_Delta}
  \Delta^k_{\ \, ij} := {1\over 2} \tgm^{kl} 
		\left( \cD_i \tgm_{lj} + \cD_j \tgm_{il}
			- \cD_l\tgm_{ij} \right) .
\ee
$\Delta^k_{\ \, ij}$ can also be viewed as the difference between
the Christoffel symbols\footnote{Recall that, while Christoffel symbols 
do not constitute the components of any tensor field, 
the difference between two sets
of them does.} of $\tna_i$ ($\tilde\Gamma^k_{\ \, ij}$)
and those of $\cD_i$ ($\bar\Gamma^k_{\ \, ij}$):
\be
	\Delta^k_{\ \, ij} = \tilde\Gamma^k_{\ \, ij} - 
	\bar\Gamma^k_{\ \, ij} 	.
\ee

The general formula for the variation of the determinant applied
to the matrix $\tgm_{ij}$ writes, once combined with 
Eq.~(\ref{e:dettgm_f}), 
\be \label{e:variation_tgm}
	\delta \ln f = \delta \ln \tgm = \tgm^{ij} \, \delta \tgm_{ij} ,
\ee
for any infinitesimal variation $\delta$ which obeys Leibniz rule. 
In the special case $\delta = \cD_k$, we deduce immediately that 
\be \label{e:variation_tgm2}
	\tgm^{ij} \cD_k \tgm_{ij} = 2 \Delta^l_{\ \, kl} = 0 .
\ee

A useful property of $\w{\tna}$ is that the divergence with respect
to it of any vector field $\w{V}$ coincides with the divergence with respect
to the flat covariant derivative $\w{\cD}$:
\be
	\tna_k V^k = \cD_k V^k \ . \label{e:div_vector}
\ee
This follows from the standard expression of the divergence 
in terms of partial derivatives with
respect to the coordinates $(x^i)$, and from Eq.~(\ref{e:dettgm_f}).

\subsection{Conformal decomposition} \label{s:conf_decomp}

We represent the traceless part of the extrinsic curvature
by
\be
	A^{ij} := \Psi^{4} \left( K^{ij} - {1\over 3} K \gm^{ij}
		\right).
\ee
Again, contrary to the $A^{ij}$ of the BSSN 
formulation \cite{ShibaN95,BaumgS99},
this quantity is a tensor field and not a tensor density. 
We introduce the following related type $\left({0\atop 2}\right)$
tensor field:
\be
	\taa_{ij} := \tgm_{ik} \tgm_{jl} A^{kl}
	= \Psi^{-4} \left( K_{ij} - {1\over 3} K \gm_{ij}
		\right),
\ee
which can be seen as $A^{ij}$ with the indices lowered by 
$\tgm_{ij}$, instead of $\gm_{ij}$. 
Both $A^{ij}$ and $\taa_{ij}$ are traceless, in the sense that
\be
	\gm_{ij} A^{ij} = \tgm_{ij} A^{ij} = 0 \qquad \mbox{and} \qquad
	\gm^{ij} \taa_{ij} = \tgm^{ij} \taa_{ij} = 0 .
\ee

The Ricci tensor $\w{R}$ of the covariant derivative $\w{D}$
(associated with the physical 3-metric $\w{\gm}$) is related to the
Ricci tensor $\w{\tilde R}$ of the covariant derivative $\w{\tna}$
(associated with the conformal metric $\w{\tgm}$) by:
\bea 
	R_{ij} & = & \tilde R_{ij} - 2 \tna_i \tna_j \Phi
	+ 4 \tna_i \Phi \, \tna_j\Phi \nonumber \\*
	& & - 2\left( \tna^k \tna_k\Phi
	 + 2 \tna_k\Phi \, \tna^k\Phi \right)\, \tgm_{ij}  , \label{e:R_tR}
\eea
where 
\be
    \Phi := \ln\Psi 
\ee 
and we have introduced the notation 
[in the same spirit as in Eq.~(\ref{e:def_upcD})]
\be \label{e:def_uptD}
	\tna^i :=\tgm^{ij} \tna_j .
\ee
The trace of Eq.~(\ref{e:R_tR}) gives
\be \label{e:trR_trtR}
	R = \Psi^{-4} \left( \tilde R 
	- 8 \tna_k \tna^k \Phi
	- 8 \tna_k \Phi \, \tna^k \Phi \right) ,
\ee
where we have introduced the scalar curvature of the metric
$\tgm_{ij}$:
\be \label{e:tildeR}
	\tilde R:=\tgm^{ij} \tilde R_{ij} .
\ee
An equivalent form of Eq.~(\ref{e:trR_trtR}) is
$R = \Psi^{-4} \tilde R - 8 \Psi^{-5} \tna_k \tna^k \Psi$,
which agrees with Eq.~(54) of York \cite{York79}.

Thanks to Eq.~(\ref{e:trR_trtR}), the Hamiltonian constraint 
(\ref{e:ham_constr0}) can be re-written
\be \label{e:ham_constr1}
   \tna_k\tna^k \Phi + \tna_k\Phi \tna^k\Phi
   = {\tilde R\over 8}  - \Psi^4 \left( 2\pi E 
   	+ {1\over 8} \taa_{kl}A^{kl} 
	- {K^2\over 12} \right) .
\ee
This equation is equivalent to Eq.~(70) of York \cite{York79}.
The momentum constraint (\ref{e:mom_constr0}) becomes
\be \label{e:mom_constr1}
	\tna_j A^{ij} + 6 A^{ij} \tna_j\Phi
		-{2\over 3} \tna^i K = 8\pi\Psi^4 J^i ,
\ee
which agrees with Eq.~(44) of Alcubierre et al.
\cite{AlcubABSS00} in the special case of Cartesian coordinates
(these Authors are using the quantity $\Phi' = \Phi + 1/12\: \ln f$,
with $f=1$ in Cartesian coordinates).

The trace of the dynamical equation (\ref{e:evol_K0}) [combined
with the Hamiltonian constraint (\ref{e:ham_constr0})] gives rise to
an evolution equation for the trace $K$ of the extrinsic curvature:
\bea 
	\der{K}{t} & - & \beta^k \tna_k K  = - \Psi^{-4} \left(
	\tna_k\tna^k N + 2 \tna_k\Phi \, \tna^k N \right) \nonumber \\*
	& & + N\left[ 4\pi (E+S) + \taa_{kl} A^{kl}
	+ {K^2\over 3} \right] ,  \label{e:evol_K1tr}
\eea
whereas the traceless part of Eq.~(\ref{e:evol_K0}) becomes
\begin{widetext}
\bea
  \der{A^{ij}}{t} & - & \pounds_{\w{\beta}} A^{ij} 
  - {2\over 3} \tna_k\beta^k \, A^{ij} = - \Psi^{-6}
   \left( \tna^i\tna^j Q - {1\over 3} \tna_k \tna^k Q \: \tgm^{ij} 
   \right) 
  + \Psi^{-4} \Bigg\{ N \left(
  \tgm^{ik} \tgm^{jl} \tilde R_{kl} + 8 \tna^i \Phi\, \tna^j \Phi
  \right) \nonumber \\
  & & + 4 \left( \tna^i\Phi \, \tna^j N + \tna^j\Phi \,
  \tna^i N \right) 
   - {1\over 3} \left[  N \left( \tilde R + 8 \tna_k\Phi \tna^k\Phi
  \right) + 8 \tna_k\Phi \tna^k N \right] \, \tgm^{ij} \Bigg\} \nonumber \\
  & & + N \left[ K A^{ij} + 2 \tgm_{kl} A^{ik} A^{jl} 
   - 8\pi \left( \Psi^4 S^{ij} - {1\over 3} S \tgm^{ij} \right) 
  \right] , \label{e:evol_K1}
\eea
\end{widetext}
 where we have introduced the scalar field
\be
	Q := \Psi^2 N .
\ee
$Q$ has the property to gather the second order derivatives of $N$
and $\Psi$ in Eq.~(\ref{e:evol_K1}). Moreover, in the stationary case, 
it has no asymptotic monopolar term (decaying like $1/r$), as discussed
in \cite{GourgGB02}. An elliptic equation for $Q$ is obtained by
combining Eqs.~(\ref{e:ham_constr1}) and (\ref{e:evol_K1tr}):
\bea
   \tna_k \tna^k Q & = & \Psi^2 \Bigg[ \Psi^4 N \left( 4\pi S
   + {3\over 4} \taa_{kl}A^{kl} +{K^2\over 2} \right)  \nonumber \\*
   & & + N \left( {1\over 4}\tilde R + 2
   	\tna_k \Phi \, \tna^k \Phi \right) 
	+ 2  \tna_k \Phi \, \tna^k N \nonumber \\*
   & & - \Psi^4 \left( \der{K}{t}
    -  \beta^k \tna_k K \right) \Bigg]  . \label{e:lapQ1}
\eea

The trace and traceless parts of the kinematical relation (\ref{e:kin0}) 
between $\w{K}$ and $\w{\gm}$ result respectively in
\be \label{e:kin1_tr}
   \der{\Psi}{t} = \beta^k\tna_k\Psi + {\Psi\over 6}
   \left( \tna_k\beta^k - N K \right) 
\ee
and
\be \label{e:kin1_st}
  \der{\tgm^{ij}}{t} - \pounds_{\w{\beta}} \tgm^{ij} 
  - {2\over 3} \tna_k\beta^k \, \tgm^{ij}
  = 2N A^{ij}  .	
\ee

\section{Einstein equations in terms of the flat covariant derivative}
\label{s:3p1flat}

It is worth to write the Einstein equations, not in terms of the 
conformal covariant derivative $\w{\tna}$, as done above, but in
terms of the flat covariant derivative $\w{\cD}$, because
(i) numerical resolution usually proceeds through linear
operators expressed in terms of $\w{\cD}$ (and deals with 
non-linearities via iterations), and (ii) the Dirac gauge we aim to use
is expressed in terms of $\w{\cD}$.

\subsection{Ricci tensor of $\w{\tna}$ in terms of the
flat derivatives of $\w{\tgm}$} \label{s:Ricci_flat_cov}

The Ricci tensor $\w{\tilde R}$ of the covariant derivative $\w{\tna}$ 
which appears in the equations of Sec.~\ref{s:conf_decomp}
can be expressed in terms of the flat covariant derivatives of the 
conformal metric $\w{\tgm}$ as
\bea
   \tilde R_{ij}  & = & - {1\over 2} \tgm^{kl} \left( 
   \cD_k \cD_l \tgm_{ij} - \cD_k \cD_i \tgm_{lj}
   - \cD_k \cD_j \tgm_{il} \right) \nonumber \\*
   & & + {1\over 2} \cD_k \tgm^{kl} \left( \cD_i \tgm_{lj} + \cD_j \tgm_{il}
   	- \cD_l \tgm_{ij} \right) \nonumber \\*
	& & - \Delta^k_{\ \, il} \Delta^l_{\ \, jk} . \label{e:ricci1}
\eea
This equation agrees with Eq.~(2.17) of 
\cite{ShibaN95}, provided it is restricted to Cartesian coordinates,
for which $\cD_i \rightarrow \partial_i$ and 
$\Delta^k_{\ \, ij} \rightarrow \tilde\Gamma^k_{\ \, ij}$.
After some manipulations, Eq.~(\ref{e:ricci1}) can be written as
\bea
   \tilde R_{ij}  & = & - {1\over 2} \Big( \tgm^{kl}
   \cD_k \cD_l \tgm_{ij} + \tgm_{ik} \cD_j H^k  + \tgm_{jk} \cD_j H^k
   \nonumber \\*
   & & + H^k \cD_k \tgm_{ij} + \cD_i \tgm^{kl} \cD_k \tgm_{lj}
   + \cD_j \tgm^{kl} \cD_k \tgm_{il} \Big) \nonumber \\*
   & & - \Delta^k_{\ \, il} \Delta^l_{\ \, jk} , \label{e:ricci2}
\eea
where we have introduced the vector field
\be \label{e:def_H}
	H^i := \cD_j \tgm^{ij} = - \tgm^{kl} \Delta^i_{\ \, kl}
\ee
[the second equality results from Eq.~(\ref{e:def_Delta})].
If we restrict ourselves to Cartesian coordinates 
($\cD_i \rightarrow \partial_i$,
$\Delta^i_{\ \, kl} \rightarrow \tilde \Gamma^i_{\ \, kl}$), 
the quantity $H^i$ coincides with minus the ``conformal connection functions''
$\tilde\Gamma^i$ introduced by Baumgarte and Shapiro \cite{BaumgS99}:
$\tilde\Gamma^i = - H^i$. 
Moreover after some rearrangements,
the expression (\ref{e:ricci2}) for the Ricci tensor can be shown to 
agree with Eq.~(22) of \cite{BaumgS99}. The motivation for the
writing (\ref{e:ricci2}) of the Ricci tensor traces back to 
Nakamura, Oohara and Kojima \cite{NakamOK87}; it consists in letting appear 
a Laplacian acting on $\tgm_{ij}$ [first term on the right-hand side
of Eq.~(\ref{e:ricci2})] and put all the other second order derivatives
of $\tgm_{ij}$ into derivatives of $H^i$. This is very similar
to the decomposition of the 4-dimensional Ricci tensor which motivates
the introduction of harmonic coordinates; note that in general the
principal part of the Ricci tensor contains 4 terms with second-order
derivatives of the metric; we have only 3 in Eq.~(\ref{e:ricci2})
because $\det\tgm_{ij} = f$.

Starting from Eq.~(\ref{e:ricci2}), we obtain, after some computations,
an expression of the Ricci tensor in terms of the flat covariant derivatives
of $\tgm^{ij}$, instead of $\tgm_{ij}$:
\begin{widetext}
\bea
  \tgm^{ik} \tgm^{jl} \tilde R_{kl} & = & {1\over 2} \Bigg(
  \tgm^{kl} \cD_k \cD_l \tgm^{ij} - \tgm^{ik} \cD_k H^j - \tgm^{jk} \cD_k H^i
  + H^k \cD_k\tgm^{ij} - \cD_l \tgm^{ik} \cD_k \tgm^{jl} 
  - \tgm_{kl} \tgm^{mn} \cD_m \tgm^{ik} \, \cD_n \tgm^{jl} \nonumber \\
   & &+ \tgm^{ik} \tgm_{ml} \cD_k \tgm^{mn} \, \cD_n \tgm^{jl} 
   + \tgm^{jl} \tgm_{kn} \cD_l \tgm^{mn} \, \cD_m \tgm^{ik} 
  +{1\over 2} \tgm^{ik} \tgm^{jl} \cD_k\tgm_{mn} \, \cD_l \tgm^{mn}
  \Bigg) \label{e:ricci3}.
\eea  
\end{widetext}
If we restrict ourselves to Cartesian coordinates,
the terms with second derivatives of $\tgm^{ij}$, i.e. the first
three terms in the above equation, agree with Eq.~(12) of 
\cite{AlcubBMS99}.

The curvature scalar  $\tilde R$ defined from the Ricci
tensor $\w{\tilde R}$ by Eq.~(\ref{e:tildeR}) is basically minus the 
flat divergence of $\w{H}$ plus some quadratic terms:
\be \label{e:tildeR_divH}
	\tilde R = - \cD_k H^k + {1\over 4} \tgm^{kl} \cD_k \tgm^{ij}
	\cD_l \tgm_{ij} - {1\over 2} \tgm^{kl} \cD_k \tgm^{ij} \cD_j \tgm_{il} . 
\ee

\subsection{Definition of the potentials $h^{ij}$}

We will numerically solve not for the conformal metric $\w{\tgm}$
but for the deviation $\w{h}$ of the inverse conformal
metric $\tgm^{ij}$ from the inverse flat metric, defined by the formula
\be \label{e:def_h}
	\tgm^{ij} =: f^{ij} + h^{ij} .
\ee
$\w{h}$ is a symmetric tensor field on $\Sigma_t$ 
of type $\left({2\atop 0}\right)$
(``twice contravariant tensor'' $h^{ij}$) and we will manipulate it
as such, without introducing any bilinear form (``twice covariant tensor''
$h_{ij}$) dual to it.

The flat covariant derivatives of $\w{h}$ coincide with those
of $\tgm^{ij}$: $\cD_k \tgm^{ij} = \cD_k h^{ij}$.
In particular the vector field $\w{H}$ introduced in Eq.~(\ref{e:def_H})
is the flat divergence of $\w{h}$:
\be 
	H^i = \cD_j h^{ij} .
\ee
Thanks to the splitting (\ref{e:def_h}), we can express the 
differential operator
$\tgm^{kl}\cD_k\cD_l$ which appears in the equations listed in 
Sec.~\ref{s:Ricci_flat_cov} as 
$\tgm^{kl}\cD_k\cD_l = \Delta + h^{kl} \cD_k\cD_l$ ,
where $\Delta$ is the Laplacian operator associated with the flat
metric:
\be \label{e:def_flat_lap}
	\Delta := f^{kl}\cD_k\cD_l = \cD_k \cD^k .
\ee 

\subsection{Einstein equations in terms of $\w{h}$ and $\w{\cD}$}

Inserting Eq.~(\ref{e:tildeR_divH}) into
the combination (\ref{e:lapQ1})
of the Hamiltonian constraint and the trace of
the spatial part of the dynamical Einstein equations 
leads to 
\begin{widetext}
\bea
  \Delta  Q &=& -h^{kl} \cD_k \cD_l Q - H^k \cD_k Q 
   + \Psi^6 \left[  N \left( 4\pi S
   + {3\over 4} \taa_{kl}A^{kl} +{K^2\over 2} \right) 
   - \der{K}{t} + \beta^k \cD_k K \right] \nonumber \\*
   && + \Psi^2 \left[ N \left( - {1\over 4} \cD_k H^k 
    + {1\over 16} \tgm^{kl} \cD_k h^{ij} \cD_l \tgm_{ij} 
   - {1\over 8} \tgm^{kl} \cD_k h^{ij} \cD_j \tgm_{il}
   + 2 \tna_k \Phi \, \tna^k \Phi   \right)
   + 2 \tna_k \Phi \, \tna^k N  \right].	
   \label{e:lapQ3}
\eea
\end{widetext}
The momentum constraint (\ref{e:mom_constr1}) writes
\be 
	\cD_j A^{ij} + \Delta^i_{\ \, kl} A^{kl}
	+ 6 A^{ij} \cD_j\Phi
		-{2\over 3} \tgm^{ij} \cD_j K = 8\pi\Psi^4 J^i ,
		\label{e:mom_constr3}
\ee
with the following expression for $\Delta^i_{\ \, kl}$,
alternative to Eq.~(\ref{e:def_Delta}):
\be
    \Delta^k_{\ \, ij}	= -{1\over 2} \left( \cD^k \tgm_{ij}
    	+ h^{kl} \cD_l \tgm_{ij} + \tgm_{il} \cD_j h^{kl}
	+  \tgm_{lj} \cD_i h^{kl} \right) .
\ee
Taking into account property~(\ref{e:div_vector}), 
the trace relation (\ref{e:kin1_tr}) can be expressed as 
\be 
   \der{\Phi}{t} - \beta^k\cD_k\Phi =  {1\over 6}
   \left( \cD_k\beta^k - N K \right) . \label{e:kin3_tr}
\ee 
The combination (\ref{e:evol_K1tr}) of the 
trace of the dynamical Einstein equations with the Hamiltonian
constraint equations becomes
\bea
	&& \der{K}{t}  - \beta^k \cD_k K  = - \Psi^{-4} \big(
	\Delta N + h^{kl} \cD_k \cD_l N + H^k \cD_k N \nonumber \\* 
	&& \ + 2 \tna_k\Phi \, \tna^k  N \big) 
        + N\left[ 4\pi (E+S) + \taa_{kl} A^{kl}
	+ {K^2\over 3} \right] . \label{e:evol_K3tr} 
\eea
After some computations, 
the traceless kinematical relation (\ref{e:kin1_st}) and the 
traceless part (\ref{e:evol_K1}) of the dynamical Einstein equations
become respectively
\begin{widetext}
\bea 
  \der{h^{ij}}{t} - \pounds_{\w{\beta}} h^{ij} 
  - {2\over 3} \cD_k\beta^k \; h^{ij} & = & 2N A^{ij} 
  - (L\beta)^{ij} , \label{e:kin3_st} \\
\der{A^{ij}}{t}  -  \pounds_{\w{\beta}} A^{ij} 
- {2\over 3} \cD_k \beta^k\; A^{ij}
   & = & {N\over 2 \Psi^4} \left( \Delta h^{ij} - \cD^i H^j
    -\cD^j H^i +{2\over 3} \cD_k H^k \, f^{ij} \right) \nonumber \\*
   & & - {1\over 2\Psi^6} \left( \cD^i h^{jk} + \cD^j h^{ik}
	- \cD^k h^{ij} - {2\over 3} H^k f^{ij} \right) \cD_k Q 
        +{\cal S}^{ij} , \label{e:evol_K3st}
\eea
where ${\cal S}^{ij}$ is given by
\bea
    {\cal S}^{ij} & := & \Psi^{-4} \Bigg\{ N \left( {\tilde R}_*^{ij}
        + 8 \tna^i\Phi \tna^j\Phi \right)
        + 4 \left( \tna^i\Phi\tna^j N + \tna^j\Phi\tna^i N \right) 
     - {1\over 3} \Big[ N\left( ({\tilde R}_* + 8\tna_k\Phi \tna^k\Phi)
     \tgm^{ij} - \cD_k H^k h^{ij} \right) \nonumber \\*
     & &   + 8 \tna_k \Phi \tna^k N \tgm^{ij}  \Big] 
        \Bigg\} 
        + N\left[ K A^{ij} + 2 \tgm_{kl} A^{ik} A^{jl}
        - 8\pi \left( \Psi^4 S^{ij} - {1\over 3} S\, \tgm^{ij} \right)
        \right] \nonumber \\*
   & &  -\Psi^{-6} \left[ \tgm^{ik}\tgm^{jl} \cD_k\cD_l Q
    + {1\over 2} \left( h^{ik} \cD_k h^{lj} + h^{kj} \cD_k h^{il}
        - h^{kl}\cD_k h^{ij} \right) \cD_l Q
        - {1\over 3} \left( \tgm^{kl} \cD_k \cD_l Q\; \tgm^{ij} 
            + H^k \cD_k Q \, h^{ij} \right) \right] , \label{e:def_Sij}
\eea
with
\bea 
   {\tilde R}_*^{ij} & := & {1\over 2} \Bigg[ h^{kl}\cD_k\cD_l h^{ij}
   - h^{ik} \cD_k H^j -  h^{jk} \cD_k H^i
    + H^k \cD_k h^{ij} - \cD_l h^{ik} \cD_k h^{jl}
    - \tgm_{kl}\tgm^{mn} \cD_m h^{ik} \cD_n h^{jl} \nonumber \\*
    & & + \tgm_{nl} \cD_k h^{mn} \left( \tgm^{ik} \cD_m h^{jl}
        + \tgm^{jk} \cD_m h^{il} \right) + {1\over 2}\tgm^{ik}\tgm^{jl}
        \cD_k h^{mn} \cD_l\tgm_{mn} \Bigg] , \label{e:def_Rijstar}  \\
  {\tilde R}_* & := &  {1\over 4} \tgm^{kl} \cD_k h^{mn}
	\cD_l \tgm_{mn} - {1\over 2} \tgm^{kl} \cD_k h^{mn} \cD_n \tgm_{ml} .
            \label{e:def_tildeRstar} 
\eea
\end{widetext}
Finally the notation $(L\beta)^{ij}$ in Eq.~(\ref{e:kin3_st})
stands for the conformal Killing operator associated
with the flat metric $\w{f}$ and applied to the vector field $\w{\beta}$:
\be \label{e:def_confKilling}
	(L\beta)^{ij} := \cD^i \beta^j + \cD^j \beta^i 
		- {2\over 3} \cD_k\beta^k \; f^{ij} .
\ee
The writing (\ref{e:evol_K3st}) with the introduction of ${\cal S}^{ij}$
by Eq.~(\ref{e:def_Sij}) is performed in order to single out the
part which is linear in the first and second derivatives of $h^{ij}$
(a term like $h^{kl}\cD_k\cD_l h^{ij}$ or 
$h^{ik} \cD_k h^{lj} \cD_l Q$ being considered as non-linear).
In particular the quantities ${\tilde R}_*^{ij}$ and ${\tilde R}_*$
arise from the decomposition of the Ricci tensor (\ref{e:ricci3})
and Ricci scalar (\ref{e:tildeR_divH}) in linear and quadratic parts:
\be
    \tgm^{ik} \tgm^{jl} \tilde R_{kl} = {1\over 2} \left( \Delta h^{ij}
        - \cD^i H^j - \cD^j H^i \right)
        + {\tilde R}_*^{ij} , 
\ee
\be
  {\tilde R} = - \cD_k H^k + {\tilde R}_* .
\ee
Consequently ${\cal S}^{ij}$ contains no linear terms in the first and 
second-order spatial derivatives of $h^{ij}$. Regarding the time derivatives
of $h^{ij}$ (encoded in $A^{ij}$), it contains only one linear term, 
in $N K A^{ij}$.
Note also that the  covariant form $\tgm_{ij}$ of the conformal metric
which appears in the
expressions of ${\tilde R}_*^{ij}$ and ${\tilde R}_*$
is the inverse of the matrix $\tgm^{ij}$, and therefore 
can be expressed as a quadratic function of $h^{ij}$, thanks to the
fact that $\tgm = f$.


\section{Maximal slicing and Dirac gauge} \label{s:Dirac}

\subsection{Definitions and discussion} \label{s:Dirac_def}

Let us now turn to the choice of coordinates, to fully specify 
the PDE system to be solved. First regarding the foliation
$\Sigma_t$, we choose {\em maximal slicing}:
\be \label{e:max_slicing}
	K = 0 . 
\ee
This well-known type of slicing has been introduced by Lichnerowicz
\cite{Lichn44} and popularized by 
York \cite{York79,SmarrY78a}. It is often disregarded in 3-D numerical relativity
because it leads to an elliptic equation for the lapse function
(cf. discussion in Sec.~\ref{s:mot_constr}). 
However it has very nice properties: beside the well-known singularity
avoidance capability \cite{SmarrY78b}, it has been shown to be well adapted to 
the propagation of gravitational waves \cite{SmarrY78a,ShibaN95}.
 
Regarding the choice of the three coordinates $(x^i)$ on each 
slice $\Sigma_t$, we consider the Dirac gauge. In Dirac's original definition
\cite{Dirac59}, it corresponds to the requirement
\be
	\der{}{x^j} \left( \gm^{1/3} \gm^{ij} \right) = 0 . 
\ee
This writing makes sense only with Cartesian type coordinates. 
In order to allow for any type of coordinates, we define
the {\em generalized Dirac gauge} as 
\be \label{e:Dirac_gen}
	\cD_j \left[ \left({\gm\over f} \right) ^{1/3} 
		\gm^{ij} \right] = 0 .
\ee
Obviously this covariant definition is made possible thanks 
to the introduction of the flat metric $\w{f}$ on $\Sigma_t$. We recognize in 
Eq.~(\ref{e:Dirac_gen}) the flat divergence of the conformal metric:
\be
	\cD_j \tgm^{ij} = 0 . \label{e:Dirac_tgm}
\ee
Since $\cD_j f^{ij} = 0$, this condition is equivalent to the vanishing
of the flat divergence of the potential $h^{ij}$:
\be \label{e:Dirac_div_h}
	\cD_j h^{ij} = 0 , 
\ee
Another equivalent definition of the Dirac gauge is requiring
that the vector $\w{H}$ vanishes [cf. Eq.~(\ref{e:def_H})]:
\be \label{e:Dirac_H}
	H^i = 0 . 
\ee
As discussed in Sec.~\ref{s:mot_constr}, the Dirac gauge has been considered
as a candidate for a radiation gauge by Smarr and York \cite{SmarrY78a}
but disregarded in profit of the {\em minimal distortion gauge} which allows for
any choice of coordinates in the initial slice. On the contrary Dirac gauge 
fully specifies (up to some boundary conditions) 
the coordinates in the slices $\Sigma_t$, including the
initial one. This property allows the search for stationary solutions
of the proposed system of equations. In particular this allows to get
quasi-stationary initial conditions for the time evolution.  
In this respect note that the numerous conformally flat initial data computed to 
date (see Ref.~\cite{BaumgS03} for a review) automatically fulfill 
Dirac gauge, since the conformal flatness of the spatial metric $\w{\gm}$ 
is equivalent to the condition $\w{h}=0$.

Another strong motivation for choosing the Dirac gauge is that it simplifies
drastically the principal linear part of the Ricci tensor $\w{\tilde R}$
associated with the conformal metric: as seen on Eq.~(\ref{e:ricci3})
or Eq.~(\ref{e:def_Rijstar}), 
this Ricci tensor, considered as a partial differential operator acting
on $\w{h}$ reduces to the elliptic term 
$\tgm^{kl} \cD_k \cD_l h^{ij}$ in that gauge. Consequently, the second order
part of the right hand side of Eq.~(\ref{e:evol_K3st}) reduces to 
a flat Laplacian $\Delta h^{ij}$. This reduction of the Ricci
tensor to a Laplacian has been the main motivation of the promotion of $\w{H}$
as an independent variable in the BSSN formulation \cite{ShibaN95,BaumgS99}.
A related property of the Dirac gauge is that thanks to it, 
the curvature scalar $\tilde R$
of the conformal metric does not contain any second order derivative
of $\tgm^{ij}$ [set $H^k=0$ in Eq.~(\ref{e:tildeR_divH})].

Note that although Dirac gauge and minimal distortion gauge differ
in the general case, both gauges result asymptotically 
in transverse-traceless (TT) coordinates (cf. Sec.~IV of Ref.~\cite{SmarrY78a}), 
which are well adapted
to the treatment of gravitational radiation. 
Both gauges are analogous to Coulomb gauge in electrodynamics.
In 1994, Nakamura \cite{Nakam94} has used a gauge, called {\em 
pseudo-minimal shear}, which is 
related to the Dirac gauge, for it writes
$\cD^j(\dert{\tgm_{ij}}{t}) = 0$, while Dirac gauge implies
$\cD_j(\dert{\tgm_{ij}}{t}) = 0$. Note however that this 
pseudo-minimal shear does not fix the coordinates on the
initial time slice, contrary to Dirac gauge: as the 
minimal distortion condition, it only rules the 
time evolution of the coordinate system. 
The exact Dirac gauge has been employed recently in two numerical studies, 
by Kawamura, Oohara and Nakamura \cite{KawamON03}, who call it {\em
the pseudo-minimal distortion condition}, and by Shibata, Uryu and
Friedman \cite{ShibaUF04}. 

Finally let us mention that Andersson and Moncrief \cite{AnderMo03}
have shown recently that the Cauchy problem for 3+1 Einstein equations
is locally strongly well posed for a coordinate system quite similar
to maximal slicing + Dirac gauge, namely
{\em constant mean curvature ($K=t$)} and {\em spatial harmonic coordinates}
($\cD_j \left[ \left({\gm / f} \right) ^{1/2}  \gm^{ij} \right] = 0$).

\subsection{Einstein equations within maximal slicing and 
Dirac gauge} \label{s:Einstein_dirac}

Thanks to the choices (\ref{e:max_slicing}) and (\ref{e:Dirac_H}),
 the combination (\ref{e:lapQ3})
of the Hamiltonian constraint and the trace of
the spatial part of the dynamical Einstein equations 
simplifies somewhat
\bea
  && \Delta  Q = -h^{kl} \cD_k \cD_l Q 
   + \Psi^6 \left[  N \left( 4\pi S
   + {3\over 4} \taa_{kl}A^{kl} \right) 
    \right] \nonumber \\*
   &&\quad + 2 \Psi^2 \left[ N \left(  {{\tilde R}_* \over 8} 
   + \tna_k \Phi \tna^k \Phi \right)
   + \tna_k \Phi \tna^k N   \right],	
   \label{e:lapQ4}
\eea
where we have let appear the quadratic quantity ${\tilde R}_*$
defined by Eq.~(\ref{e:def_tildeRstar}). 
Note that thanks to Dirac gauge, the right hand side of the
above  equation
does not contain any second order derivative of $h^{ij}$. 

The momentum constraint (\ref{e:mom_constr3}) becomes
\be 
	\cD_j A^{ij} + \Delta^i_{\ \, kl} A^{kl}
	+ 6 A^{ij} \cD_j\Phi = 8\pi\Psi^4 J^i .
		\label{e:mom_constr4}
\ee
Now, taking the (flat) divergence of Eq.~(\ref{e:kin3_st}) and using the
fact that $\partial/\partial t$ commutes with $\cD_i$, thanks to 
property (\ref{e:f_notime}), the Dirac gauge leads to an expression of
the divergence of $A^{ij}$ which does not contain any time derivative
of the shift vector nor any second derivative of $h^{ij}$: 
\bea
   \cD_j A^{ij} & = & - {A^{ij}\over N} \cD_j N 
   + {1\over 2N} \Bigg[ \Delta \beta^i + 
   {1\over 3} \cD^i \left( \cD_j\beta^j \right) \nonumber \\*
   && + h^{kl} \cD_k \cD_l \beta^i + {1\over 3} h^{ik} \cD_k
   \left( \cD_l \beta^l \right) \Bigg] .
\eea
Inserting this relation into the momentum constraint equation
(\ref{e:mom_constr4})
results in an elliptic equation for $\w{\beta}$:
\bea 
  \Delta\beta^i & + & {1\over 3} \cD^i\left(\cD_j\beta^j\right) = 
   16\pi N \Psi^4 J^i + 2 A^{ij} \cD_j N \nonumber \\*
   &&  - 12 N A^{ij} \cD_j\Phi - 2 N \Delta^i_{\ \, kl} A^{kl} \nonumber \\*
   && - h^{kl} \cD_k \cD_l \beta^i - {1\over 3} h^{ik} \cD_k \cD_l \beta^l .
        \label{e:Poisson_beta}
\eea

Thanks to maximal slicing, the kinematical trace relation (\ref{e:kin3_tr}) 
reduces to 
\be 
   \der{\Phi}{t} - \beta^k\cD_k\Phi =  {1\over 6} \cD_k\beta^k . 
                                \label{e:kin4_tr}
\ee 
The combination (\ref{e:evol_K3tr}) of the 
trace of the dynamical Einstein equations with the Hamiltonian
constraint equations becomes an elliptic equation for the lapse
function:
\bea 
	\Delta N & = & \Psi^4 N\left[ 4\pi (E+S) + \taa_{kl} A^{kl}
	\right] 
	 - h^{kl} \cD_k \cD_l N \nonumber \\*
	&& - 2 \tna_k\Phi \, \tna^k  N  . \label{e:evol_K4tr} 
\eea

In Dirac gauge + maximal slicing, the time evolution system 
(\ref{e:kin3_st})-(\ref{e:evol_K3st}) becomes
\bea 
  \der{h^{ij}}{t} &- &\pounds_{\w{\beta}} h^{ij} 
  - {2\over 3} \cD_k\beta^k \; h^{ij} =  2N A^{ij} 
  - (L\beta)^{ij}   \label{e:kin4_st} \\
\der{A^{ij}}{t} & - & \pounds_{\w{\beta}} A^{ij} 
- {2\over 3} \cD_k \beta^k\; A^{ij}
   = {N\over 2 \Psi^4} \Delta h^{ij} +{\cal S}^{ij}  \nonumber \\*
    && - {1\over 2\Psi^6} \left( \cD^i h^{jk} + \cD^j h^{ik} 
	 - \cD^k h^{ij}  \right) \cD_k Q 
        , \label{e:evol_K4st}
\eea
where ${\cal S}^{ij}$ is slightly simplified to 
\begin{widetext}
\bea
    {\cal S}^{ij} & = & \Psi^{-4} \Bigg\{ N \left( {\tilde R}_*^{ij}
        + 8 \tna^i\Phi \tna^j\Phi \right)
        + 4 \left( \tna^i\Phi\tna^j N + \tna^j\Phi\tna^i N \right) 
     - {1\over 3} \Big[ N\left( ({\tilde R}_* + 8\tna_k\Phi \tna^k\Phi)
     \tgm^{ij} \right) \nonumber \\*
     & &   + 8 \tna_k \Phi \tna^k N \, \tgm^{ij}  \Big] 
        \Bigg\} 
        + 2 N\left[  \tgm_{kl} A^{ik} A^{jl}
        - 4\pi \left( \Psi^4 S^{ij} - {1\over 3} S\, \tgm^{ij} \right)
        \right] \nonumber \\*
   & &  -\Psi^{-6} \left[ \tgm^{ik}\tgm^{jl} \cD_k\cD_l Q
    + {1\over 2} \left( h^{ik} \cD_k h^{lj} + h^{kj} \cD_k h^{il}
        - h^{kl}\cD_k h^{ij} \right) \cD_l Q
        - {1\over 3} \left( \tgm^{kl} \cD_k \cD_l Q\; \tgm^{ij} 
             \right) \right] , \label{e:Sij_Dirac}
\eea
with
\bea 
   {\tilde R}_*^{ij} & = & {1\over 2} \Bigg[ h^{kl}\cD_k\cD_l h^{ij}
    - \cD_l h^{ik} \cD_k h^{jl}
    - \tgm_{kl}\tgm^{mn} \cD_m h^{ik} \cD_n h^{jl} 
     + \tgm_{nl} \cD_k h^{mn} \left( \tgm^{ik} \cD_m h^{jl}
        + \tgm^{jk} \cD_m h^{il} \right) \nonumber \\*
        && + {1\over 2}\tgm^{ik}\tgm^{jl}
        \cD_k h^{mn} \cD_l\tgm_{mn} \Bigg] .  \label{e:Rijstar_Dirac} 
\eea
\end{widetext}
The quadratic term ${\tilde R}_*$ in Eq.~(\ref{e:Sij_Dirac})
is unchanged and is given
by Eq.~(\ref{e:def_tildeRstar}).
The Lie derivatives along the shift vector field $\w{\beta}$
which appear in Eqs.~(\ref{e:kin4_st}) and (\ref{e:evol_K4st}) 
can be expressed in terms of the flat covariant derivative 
$\w{\cD}$ by the standard formula:
\bea
    \pounds_{\w{\beta}} h^{ij} & = & \beta^k \cD_k h^{ij}
    	- h^{kj} \cD_k \beta^i -  h^{ik} \cD_k \beta^j , \\
    \pounds_{\w{\beta}} A^{ij} & = & \beta^k \cD_k A^{ij}
    	- A^{kj} \cD_k \beta^i -  A^{ik} \cD_k \beta^j .
\eea

\subsection{Wave equation for $h^{ij}$} \label{s:wave_R3}

As discussed in Sec.~\ref{s:Dirac_def}, one of the main motivations 
for using Dirac gauge is that it changes the second order operator
acting on $h^{ij}$ in Eq.~(\ref{e:evol_K4st}) to a mere Laplacian. 
It is therefore tempting to write the first order time evolution system
(\ref{e:kin4_st})-(\ref{e:evol_K4st}) as a (second order) wave equation
for $h^{ij}$.
Note that the first order operator $\partial/\partial t - \pounds_{\w{\beta}}$
which appear on the l.h.s. of the system (\ref{e:kin4_st})-(\ref{e:evol_K4st})
is nothing but the Lie derivative along the vector $N\w{n}$.
Its square is
\be
    \left( \der{}{t} - \pounds_{\w{\beta}} \right) ^2 h^{ij} =
        \dder{h^{ij}}{t} - 2\pounds_{\w{\beta}} \der{h^{ij}}{t}
        + \pounds_{\w{\beta}}\pounds_{\w{\beta}} h^{ij}
            - \pounds_{\w{\dot\beta}} h^{ij} ,  \label{e:der_sec_h}
\ee
with the short-hand notation 
\be
    \dot\beta^i := \der{\beta^i}{t} . 
\ee
Applying the operator $\partial/\partial t - \pounds_{\w{\beta}}$
to Eq.~(\ref{e:kin4_st}) and inserting Eqs.~(\ref{e:der_sec_h})
and (\ref{e:evol_K4st}) in the result leads to the wave equation
\begin{widetext}
\bea
    \dder{h^{ij}}{t} & - & {N^2\over \psi^4} \Delta h^{ij}
    - 2 \pounds_{\w{\beta}} \der{h^{ij}}{t} 
    + \pounds_{\w{\beta}}\pounds_{\w{\beta}} h^{ij} = 
    \pounds_{\w{\dot\beta}} h^{ij} + {4\over 3} \cD_k\beta^k 
    \left( \der{}{t} - \pounds_{\w{\beta}} \right) h^{ij}
    - {N\over \Psi^6} \cD_k Q \left( \cD^i h^{jk} + \cD^j h^{ik} 
	 - \cD^k h^{ij}  \right)  \nonumber \\*
&& + {1\over N} \left[ \left( \der{}{t} - \pounds_{\w{\beta}} \right) N
    \right] \left[ \left( \der{}{t} - \pounds_{\w{\beta}} \right) h^{ij}
     - {2\over 3} \cD_k \beta^k h^{ij} + (L\beta)^{ij} \right]
     + {2\over 3} \left[ \left( \der{}{t} - \pounds_{\w{\beta}} \right)
        \cD_k\beta^k - {2\over 3} (\cD_k\beta^k)^2 \right] h^{ij}
                \nonumber  \\*
 && + 2N {\cal S}^{ij} - \left( \der{}{t} - \pounds_{\w{\beta}} \right) (L\beta)^{ij}
    + {2\over 3} \cD_k\beta^k (L\beta)^{ij} .           \label{e:wave_hij}
\eea
\end{widetext}
Note that the left-hand side of the above equation contains all the
second-order derivatives (both in time and space) of $h^{ij}$, at the 
linear order. Actually the only second-order derivative of $h^{ij}$
on the right-hand side is the non-linear term 
$h^{kl}\cD_k\cD_l h^{ij}$ contained
in ${\cal S}^{ij}$ via ${\tilde R}^{ij}_*$
[cf. Eqs.~(\ref{e:Sij_Dirac}) and (\ref{e:Rijstar_Dirac})]. 

Let us rewrite Eq.~(\ref{e:wave_hij}) as a flat-space tensorial wave equation:
\be
    \square h^{ij} = \sigma^{ij} + (L\dot\beta)^{ij} , \label{e:dalembert_hij}
\ee
where $\square$ denotes the d'Alembert operator associated with the
flat metric $\w{f}$ [cf. Eq.~(\ref{e:def_flat_lap})]:
\be
    \square := - \dder{}{t} + \Delta
\ee
and $\sigma^{ij}$ is given by
\begin{widetext}
\bea
    \sigma^{ij} & := & \left(1 - {N^2\over \psi^4} \right) \Delta h^{ij}
    - 2 \pounds_{\w{\beta}} \der{h^{ij}}{t} 
    + \pounds_{\w{\beta}}\pounds_{\w{\beta}} h^{ij} 
    - \pounds_{\w{\dot\beta}} h^{ij} - {4\over 3} \cD_k\beta^k 
    \left( \der{}{t} - \pounds_{\w{\beta}} \right) h^{ij}
    + {N\over \Psi^6} \cD_k Q \left( \cD^i h^{jk} + \cD^j h^{ik} 
	 - \cD^k h^{ij}  \right)  \nonumber \\*
&& - {1\over N} \left[ \left( \der{}{t} - \pounds_{\w{\beta}} \right) N
    \right] \left[ \left( \der{}{t} - \pounds_{\w{\beta}} \right) h^{ij}
     - {2\over 3} \cD_k \beta^k h^{ij} + (L\beta)^{ij} \right]
     - {2\over 3} \left[ \left( \der{}{t} - \pounds_{\w{\beta}} \right)
        \cD_k\beta^k - {2\over 3} (\cD_k\beta^k)^2 \right] h^{ij}
                \nonumber  \\*
 && - 2N {\cal S}^{ij}  - \pounds_{\w{\beta}} (L\beta)^{ij}
    - {2\over 3} \cD_k\beta^k (L\beta)^{ij} .           \label{e:sigma_ij}
\eea
\end{widetext}
Note that we have not included into $\sigma^{ij}$ the 
term\footnote{Eq.~(\ref{e:dsdtLbeta}) holds thanks to the property
(\ref{e:f_notime}).} 
\be
    \der{}{t} (L\beta)^{ij} = (L\dot\beta)^{ij} \label{e:dsdtLbeta}
\ee
which appears in the right-hand side of Eq.~(\ref{e:wave_hij}).
Consequently this term appears explicitly in the right-hand side of
Eq.~(\ref{e:dalembert_hij}). 

At a given time step during the evolution,
$\sigma^{ij}$ is considered as a fixed source in Eq.~(\ref{e:dalembert_hij}), 
so that the problem is reduced to solving a flat space
wave equation. 
Since $\w{\cD}$ and $\square$ commute (thanks to the time-independence
of $\w{f}$), the source  $\sigma^{ij} + (L\dot\beta)^{ij} $ must be 
divergence-free in order for the solution $h^{ij}$ of Eq.~(\ref{e:dalembert_hij})
to satisfy Dirac gauge (\ref{e:Dirac_div_h}). This means that one must have
\be
    \cD_j (L\dot\beta)^{ij} = - \cD_j \sigma^{ij} , 
\ee
or, from the definition (\ref{e:def_confKilling}) of the conformal Killing
operator and the vanishing of $\w{f}$'s Riemann tensor, 
\be
   \Delta{\dot\beta}^i + {1\over 3} \cD^i\left(\cD_j{\dot\beta}^j\right)
    = - \cD_j \sigma^{ij} . 
\ee
The above elliptic equation fully determines $\w{\dot\beta}$
(up to some boundary conditions), and therefore, by direct
time integration, $\w{\beta}$. This shows clearly
that the shift vector propagates the Dirac spatial coordinates $(x^i)$ from
one slice $\Sigma_t$ to the next one. Hence we recover the traditional
interpretation of the shift vector. 
On the other side, $\w{\beta}$ can be computed from the combination 
(\ref{e:Poisson_beta})
of the momentum constraint and Dirac gauge condition. Both ways must 
yield the same result. However, from the numerical point of view,
they may not be equivalent (due to numerical errors) and a strategy
to compute the best value of $\w{\beta}$ must be devised.

Note that, since we reduce the time evolution problem to a second-order wave 
equation for $h^{ij}$, at each step, the extrinsic curvature term
$A^{ij}$ must be deduced from the time derivative of $h^{ij}$ and the
spatial derivatives of the shift vector by inverting Eq.~(\ref{e:kin4_st}):
\be
    A^{ij} = {1\over 2N} \left[ (L\beta)^{ij}
        + \der{h^{ij}}{t} - \pounds_{\w{\beta}} h^{ij}
        - {2\over 3} \cD_k\beta^k \, h^{ij} \right] . \label{e:Aij_calcul}
\ee

\subsection{Transverse traceless decomposition}\label{s:decomp_TT} 

The generalized Dirac gauge, expressed as Eq.~(\ref{e:Dirac_div_h}),
makes the potential $\w{h}$ a transverse tensor field with respect
to the metric $\w{f}$. However, the trace of $\w{h}$ with respect to the 
metric $\w{f}$,
\be \label{e:def_trace_h}
	h := f_{ij} h^{ij}, 
\ee
does not vanish in general, except in the linearized approximation.
Therefore $\w{h}$ is not a transverse and traceless (TT) tensor field.
Since this latter property would considerably help the treatment of 
the wave equation, we perform a TT decomposition of $\w{h}$
according to (see e.g. Sec.~7-4.2 of ADM \cite{ArnowDM62})
\be \label{e:h_decompTT}
	h^{ij} =: {\bar h}^{ij} + {1\over 2} \left( h \, f^{ij}
		- \cD^i \cD^j \phi \right) ,
\ee
where $\phi$ is a solution of the Poisson equation
\be \label{e:Lap_Phi_h}
	\Delta \phi = h 
\ee
satisfying $\phi=0$ at spatial infinity. 
Then the trace of the term 
${1/2} \left( h \, f^{ij}- \cD^i \cD^j \phi \right)$ on the right-hand
side of Eq.~(\ref{e:h_decompTT}) is equal to $h$.
Moreover this term is divergence-free. We conclude that if $\w{h}$
is transverse (Dirac gauge), then $\w{\bar h}$
defined by Eq.~(\ref{e:h_decompTT}) is a TT
tensor\footnote{If we had removed the trace of $\w{h}$
in the ``standard'' way, by defining ${\tilde h}^{ij} := h^{ij}
- {1\over 3} h f^{ij}$, the
traceless part would not have been transverse.}:
\be \label{e:hij_TT}
	\cD_j {\bar h}^{ij} = 0  \qquad \mbox{and}  \qquad
	f_{ij} {\bar h}^{ij} = 0 .
\ee

We then decompose Eq.~(\ref{e:dalembert_hij}) into a trace part, and 
a traceless one, to get
\be \label{e:dalembert_h}
	\square h = \sigma,
\ee
\be \label{e:dalembert_hij_TT}
	\square {\bar h}^{ij} = {\bar\sigma}^{ij} + (L\dot\beta)^{ij},
\ee
where $\sigma := f_{ij} \sigma^{ij}$ and 
${\bar\sigma}^{ij}$ is the traceless part of $\sigma^{ij}$ given by 
the decomposition analogous to (\ref{e:h_decompTT}):
\be \label{e:decomp_Sij}
	\sigma^{ij} =: {\bar\sigma}^{ij} + {1\over 2} \left( \sigma \, f^{ij}
		- \cD^i \cD^j \Upsilon \right) ,
\ee
with $\Delta \Upsilon = \sigma$. Note that the quantity 
$(L\dot\beta)^{ij}$
is trace-free by the very definition of operator $L$ 
[Eq.~(\ref{e:def_confKilling})].

The search for the potentials $h^{ij}$ can then proceed along
the following steps: compute the trace $\sigma$ of the effective source 
$\sigma^{ij}$ [Eq.~(\ref{e:sigma_ij})] and
solve the Poisson equation
\be
	\Delta \Upsilon = \sigma , 
\ee
with the boundary condition $\Upsilon=0$ at spatial infinity. 
This leads to a regular solution for $\Upsilon$ because $\sigma$ 
is a fast decaying source, due to the fact that Eq.~(\ref{e:dalembert_hij})
is the traceless part, with respect to the metric $\w{\tgm}$, 
of the dynamical Einstein equations and that $\w{\tgm}\sim \w{f}$
asymptotically. The next step is to insert $\Upsilon$ and $\sigma$ into 
Eq.~(\ref{e:decomp_Sij}) to compute ${\bar\sigma}^{ij}$. 
Then one has to solve the TT wave equation 
(\ref{e:dalembert_hij_TT}) for ${\bar h}^{ij}$. A resolution technique based
on spherical coordinates and spherical tensor components will be 
presented in Sec.~\ref{s:resol_tensor_wave}. Using this technique, the 
resolution of Eq.~(\ref{e:dalembert_hij_TT}) is reduced to the resolution
of two scalar d'Alembert equations. Then one may solve the 
scalar d'Alembert equation 
\be \label{e:Box_Phi_Upsilon}
	\square \phi = \Upsilon
\ee  
for $\phi$ and compute the trace $h$ not by solving the d'Alembert equation
(\ref{e:dalembert_h}) but directly as the Laplacian of $\phi$
[cf. Eq.~(\ref{e:Lap_Phi_h})]. Inserting $h$ and $\phi$ into 
Eq.~(\ref{e:h_decompTT}) leads then to $h^{ij}$. 
An alternative approach to get $h$ will be 
discussed in Sec.~\ref{s:unit_determ}. 
It is algebraic [thus does not require to solve any d'Alembert equation
like (\ref{e:dalembert_h}) or (\ref{e:Box_Phi_Upsilon})]
and has the advantage to enforce the condition on the determinant of
$\tgm^{ij}$ [Eq.~(\ref{e:dettgm_f})].


\section{A resolution scheme based on spherical coordinates} 
\label{s:spher}

As discussed in Sec.~\ref{s:motiv_sphere}, spherical coordinates have
many advantages when treating neutron star or black hole spacetimes. 
Moreover, as we shall see below, the use of tensor components with 
respect to a spherical basis allow to compute three of the
metric components $\tgm^{ij}$ directly from the Dirac gauge condition 
(\ref{e:Dirac_tgm}).  
In this section we therefore specialize the coordinates $(x^i)$
on each hypersurface $\Sigma_t$ to spherical ones. 
Moreover we expand all the tensor fields onto a spherical basis
which is orthonormal with respect to the flat metric.

\subsection{Spherical orthonormal basis}

We introduce on $\Sigma_t$ a coordinate system $x^i=(r,\th,\ph)$
of spherical type, i.e. $r$ spans the range $[0,+\infty)$, 
$\th$ the range $[0,\pi]$ (co-latitude angle), $\ph$ 
the range $[0,2\pi)$ (azimuthal angle) and the components of 
the flat metric $\w{f}$ with respect to these coordinates are
\be
	f_{ij} = {\rm diag}\, (1,\; r^2,\; r^2\sin^2\th) .
\ee
The determinant $f$ [Eq.~(\ref{e:def_detf})] is then $f = r^4 \sin^2\th$.

From the natural vector basis associated with the coordinates $(r,\th,\ph)$, 
$\left( \dert{}{x^i} \right) = \left( \dert{}{r},\dert{}{\th},
\dert{}{\ph} \right)$, we construct the following vector fields:
\be \label{e:e_hr}
	\w{e}_r  :=  \der{}{r} , \quad
	\w{e}_\th  :=  {1\over r} \der{}{\th} ,\quad 
	\w{e}_\ph  :=  {1\over r\sin\th} \der{}{\ph} . 
\ee
$(\w{e}_{\hat i})= (\w{e}_r, \w{e}_\th, \w{e}_\ph)$ forms a basis
of the vector space tangent to $\Sigma_t$. Moreover, this basis
is orthonormal with respect to the flat metric $\w{f}$:
$f_{\hat i\hat j} ={\rm diag}(1,1,1)$. Notice that we are denoting with 
a hat the generic indices ${\hat i},{\hat j},...$ associated with this
basis, but we denote by $r,\th,\ph$ (without a hat) indices  
of specific components on this basis. 

Given a tensor field $\w{T}$ of type $\left({p\atop q}\right)$, 
the components of the 
covariant derivative $\w{\cD} \w{T}$ in the orthonormal
basis $\w{e}_{{\hat i}_1} \otimes \cdots
\otimes\w{e}_{{\hat i}_p}\otimes \cdots \otimes \w{e}^{{\hat j}_1}
\otimes \cdots \otimes \w{e}^{{\hat j}_q}\otimes \w{e}^{\hat k} $ are given by
\bea 
 \cD_{\hat k} 
 T^{{\hat i}_1\ldots {\hat i}_p}_{\quad \quad {\hat j}_1\ldots {\hat j}_q}
     & = & e_{\hat k}^{\ \, l} \der{}{x^l} 
	T^{{\hat i}_1\ldots {\hat i}_p}_{\quad \quad {\hat j}_1\ldots {\hat j}_q}
	\nonumber \\*
&& +  \sum_{r=1}^p \hat\Gamma^{{\hat i}_r}_{\ \, \hat l\hat k} \, 
T^{{\hat i}_1\ldots \hat l \ldots {\hat i}_p}_{\quad\quad \quad {\hat j}_1\ldots {\hat j}_q}
	\nonumber \\*
&& - \sum_{r=1}^q \hat\Gamma^{\hat l}_{\ \, {\hat j}_r\hat k } \, 
T^{{\hat i}_1\ldots {\hat i}_p}_{\quad \quad {\hat j}_1\ldots \hat l \ldots 
{\hat j}_q} , \label{e:cDT_ortho}
\eea
where $e_{\hat k}^{\ \, l} := {\rm diag} (1, 1/r, 1/(r\sin\th))$ is
the change-of-basis matrix defined by Eq.~(\ref{e:e_hr}), and 
the $\hat \Gamma^{\hat k}_{\ \, \hat i\hat j}$ are the connection coefficients
of $\w{\cD}$ associated with the orthonormal frame $(\w{e}_{\hat i})$; these
coefficients all vanish, except for 
\bea
  && \hat\Gamma^{r}_{\ \, \th\th} = - \hat\Gamma^{\th}_{\ \, r \th}
    = - r^{-1} \ , \qquad
    \hat\Gamma^{r}_{\ \, \ph\ph} = - \hat\Gamma^{\ph}_{\ \,  r \ph}
    = - r^{-1} \ , \nonumber \\ 
 && \hat\Gamma^{\th}_{\ \, \ph\ph} = - \hat\Gamma^{\ph}_{\ \, \th\ph}
    = - (r \tan\th)^{-1} .  \label{e:Ricci_rot_spher}
\eea

\subsection{Resolution of elliptic equations}

\subsubsection{Scalar Poisson equations} \label{s:scalar_poisson}

We have to solve two scalar elliptic equations: the Hamiltonian constraint
(combined with the trace of the dynamical Einstein equations)
Eq.~(\ref{e:lapQ4}) for $Q$ and the maximal slicing equation 
(\ref{e:evol_K4tr}) for $N$. Both equations are not strictly
Poisson equations since they contain $Q$ and $N$ on their right-hand side.
Moreover the right-hand side of Eq.~(\ref{e:lapQ4}) is non-linear in 
$Q$ (through $\Phi = (\ln N - \ln Q)/2$). Therefore these equations must
be solved by iterations, solving for a Poisson equation at each step.
Since we are using spherical coordinates, it is natural to perform
an expansion on spherical harmonics $Y_\ell^m(\th,\ph)$. The resolution 
of the scalar Poisson equation is then reduced to the resolution of a
system of second order ordinary differential equations in $r$ for each
couple $(\ell,m)$. We refer the reader to Ref.~\cite{GrandBGM01} for
further details. 

\subsubsection{Vector elliptic equation for the shift}
\label{s:vector_shift}

As we have seen in Sec.~\ref{s:Einstein_dirac}, 
the Dirac gauge condition once inserted into the
momentum constraint equation gives rise to the elliptic equation 
(\ref{e:Poisson_beta}). Using the derivation formula (\ref{e:cDT_ortho}) 
with the explicit values (\ref{e:Ricci_rot_spher}) of the connection
coefficients, we obtain the following components of this equation
with respect to the orthonormal frame $(\w{e}_{\hat i})$:
\begin{widetext}
\bea
    && \dder{\beta^r}{r} + {2\over r}\der{\beta^r}{r}
    +{1\over r^2} \left( \Delta_{\th\ph} \beta^r - 2 \beta^r 
    - 2 \der{\beta^\th}{\th} - 2{\beta^\th\over\tan\th} - {2\over\sin\th}
    \der{\beta^\ph}{\ph} \right) + {1\over 3} \der{\Theta}{r} = S(\w{\beta})^r 
                    \label{e:poisson_beta_r} \\
  && \dder{\beta^\th}{r} + {2\over r} \der{\beta^\th}{r}
  +{1\over r^2} \left( \Delta_{\th\ph} \beta^\th + 2 \der{\beta^r}{\th}
    - {\beta^\th\over\sin^2\th} 
    - 2{\cos\th\over\sin^2\th}\der{\beta^\ph}{\ph} \right)
    + {1\over 3r} \der{\Theta}{\th} = S(\w{\beta})^\th 
                                    \label{e:poisson_beta_t} \\
  && \dder{\beta^\ph}{r} + + {2\over r} \der{\beta^\ph}{r}
  +{1\over r^2} \left( \Delta_{\th\ph} \beta^\ph 
  + {2\over\sin\th} \der{\beta^r}{\ph}
   +2{\cos\th\over\sin^2\th} \der{\beta^\th}{\ph} 
   - {\beta^\ph\over\sin^2\th} \right) + {1\over 3r\sin\th} 
   \der{\Theta}{\ph} = S(\w{\beta})^\ph ,       \label{e:poisson_beta_p}
\eea
\end{widetext}
where $\Delta_{\th\ph}$ denotes the angular Laplacian:
\be 
	\Delta_{\th\ph} := \dder{}{\th} +{1\over\tan\th} \der{}{\th}
	+ {1\over\sin^2\th}\dder{}{\ph} ,  \label{e:angu_Lap}
\ee
$S(\w{\beta})^{\hat i}$ stands for the right-hand side of 
Eq.~(\ref{e:Poisson_beta}) 
and $\Theta:= \cD_k \beta^k$ is the divergence of $\w{\beta}$ with respect 
to the  flat connection $\w{\cD}$. In terms of the components with respect to 
the orthonormal frame $(\w{e}_{\hat i})$, it reads
\be
    \Theta  
    = \der{\beta^r}{r} + {2\beta^r\over r}
    + {1\over r} \left( \der{\beta^\th}{\th} + {\beta^\th\over\tan\th}
   + {1\over\sin\th} \der{\beta^\ph}{\ph} \right) . \label{e:Theta_div}
\ee
As for the scalar elliptic equations for $Q$ and $N$ discussed above,
the right-hand side $S(\w{\beta})^{\hat i}$ of 
Eqs.~(\ref{e:poisson_beta_r})-(\ref{e:poisson_beta_p}) depend (linearly) on 
$\w{\beta}$, both explicitly and via $A^{ij}$ [see Eqs.~(\ref{e:Poisson_beta})
and (\ref{e:Aij_calcul})]. Thus an iterative resolution must be
contemplated. 

Equations~(\ref{e:poisson_beta_r})-(\ref{e:poisson_beta_p}) constitute
a coupled system, since each equation contains all the components of
$\w{\beta}$. However, we can decouple the system by proceedings
as follows. First, taking the (flat) divergence of this vector system, 
and taking into account that $\w{\cD}$ and $\Delta$ commute (flat metric),
we get a scalar Poisson equation for $\Theta$ only:
\be
    \Delta\Theta = {3\over 4} \, \cD_{\hat k} S(\w{\beta})^{\hat k} . 
                            \label{e:Poisson_Theta_bet}
\ee 
Assuming this equation is solved for $\Theta$, we 
use Eq.~(\ref{e:Theta_div}) to replace the terms containing angular
components in Eq.~(\ref{e:poisson_beta_r}) to get a decoupled equation
for $\beta^r$:
\bea
    \dder{\beta^r}{r} & + & {4\over r}\der{\beta^r}{r}
        + {2\beta^r\over r^2} + {1\over r^2} \Delta_{\th\ph}\beta^r
       =  \nonumber \\*
    && S(\w{\beta})^r - {1\over 3} \der{\Theta}{r} 
        + {2\over r}\Theta .    \label{e:beta_r_4}
\eea
This equation can be solved by expanding $\beta^r$ in spherical
harmonics. An alternative approach is to set
\be 
    \chi := r \beta^r   \label{e:def_chi_beta}
\ee 
which reduces Eq.~(\ref{e:beta_r_4}) to an ordinary Poisson equation:
\be
   \Delta \chi = r S(\w{\beta})^r - {r\over 3} \der{\Theta}{r} 
        + 2 \Theta .    \label{e:Poisson_chi_bet}
\ee
This is not surprising since $\chi$ is actually a scalar field on
$\Sigma_t$: $\chi = f_{ij} r^i \beta^j$, where $\w{r}$ denotes the
``position'' vector field:
\be
    \w{r} := r \, \w{e}_r = x\, \w{e}_x + y\, \w{e}_y + z\,\w{e}_z ,
                    \label{e:r_def}
\ee
$(x,y,z)$ and $(\w{e}_x,\w{e}_y,\w{e}_z)$ being respectively the
Cartesian coordinates and Cartesian frame canonically associated with
the spherical coordinates $(r,\th,\ph)$.
Indeed, contrary to $\w{e}_r$, which is singular at the 
origin $r=0$, $\w{r}$ is a regular\footnote{As in Ref.~\cite{BardeP83},
we define a {\em regular} tensor field as a tensor field whose components
with respect to the Cartesian frame $(\w{e}_x,\w{e}_y,\w{e}_z)$
are expandable in power series of $x$, $y$ and $z$.}
vector field [this is obvious from the
second equality in Eq.~(\ref{e:r_def})]. Being the scalar
product of $\w{\beta}$ and $\w{r}$ (with respect to $\w{f}$), $\chi$ is
then a regular scalar.

Let us now discuss the resolution of the angular part.
We introduce a poloidal potential $\eta$ and a toroidal potential $\mu$
such that $\w{\beta}$ is expanded as (see also \S~13.1 of 
Ref.~\cite{MorseF53} and \S~A.2.a of Ref.~\cite{VillaB02}):
\be
    \w{\beta} = \beta^r \w{e}_r + \left[ r \w{\cD} \eta - (\w{e}_r \cdot
    \w{\cD}\eta)\, \w{r} \right] + \w{r} \times \w{\cD}\mu ,
                        \label{e:beta_eta_mu}
\ee
where the scalar product and the vectorial product are taken with respect
to the flat metric $\w{f}$. Note that the terms containing $\eta$ and $\mu$
are by construction tangent to the sphere $r={\rm const}$ and that
$\w{r} \times \w{\cD}\mu$ is
nothing but the angular momentum operator of Quantum Mechanics applied to
$\mu$. An alternative expression is 
$\w{r} \times \w{\cD}\mu = - \w{\cD}\times(\mu\, \w{r})$.
In term of components, Eq.~(\ref{e:beta_eta_mu}) results in 
\bea
    \beta^\th & = & \der{\eta}{\th} - {1\over\sin\th} \der{\mu}{\ph} 
                        \label{e:beta_th} \\
    \beta^\ph & = & {1\over\sin\th} \der{\eta}{\ph} + \der{\mu}{\th} . 
                    \label{e:beta_ph}    
\eea
It can be shown easily that the potentials $\eta$
and $\mu$ obey to the following relations:
\bea
    \Delta_{\th\ph} \eta & = & r \Theta - r \der{\beta^r}{r} - 2\beta^r 
                                        \label{e:lapang_eta_bet} \\
    \Delta_{\th\ph} \mu & = & \w{r} \cdot (\w{\cD}\times \w{\beta})
                \nonumber \\* 
       & = & \der{\beta^\ph}{\th} + {\beta^\ph\over\tan\th}
            - {1\over\sin\th}\der{\beta^\th}{\ph} . \label{e:lapang_mu_bet}
\eea
These formulas show that $\eta$ and $\mu$ are uniquely defined 
(up to the addition of some function of $r$). $\Theta$, $\beta^r=\chi/r$ and
the scalar $ \w{r} \cdot (\w{\cD}\times \w{\beta})$ being expandable
in (scalar) spherical harmonics, Eqs.~(\ref{e:lapang_eta_bet}) and 
(\ref{e:lapang_mu_bet}) show also that $\eta$ and $\mu$ are expandable
in spherical harmonics $Y_\ell^m(\th,\ph)$. The computation of $\eta$ and
$\mu$ from the components $(\beta^r,\beta^\th,\beta^\ph)$ 
can then be performed from 
Eqs.~(\ref{e:lapang_eta_bet})-(\ref{e:lapang_mu_bet}) by a mere division 
by $-\ell(\ell+1)$ (eigenvalue of the operator $\Delta_{\th\ph}$ 
corresponding to the eigenfunction $Y_\ell^m(\th,\ph)$). In the following
we call this type of computation a {\em quasi-algebraic} one. 

By a straightforward computation, it can be shown that the part
(\ref{e:poisson_beta_t})-(\ref{e:poisson_beta_p}) of the original system
is equivalent to the two Poisson equations
\bea
    \Delta\eta & = & \eta_S - {2\beta^r\over r^2} 
        - {1\over 3}\, {\Theta\over r}      \label{e:Poisson_eta_bet} \\
    \Delta\mu & = & \mu_S ,                     \label{e:Poisson_mu_bet}
\eea
where $\eta_S$ and $\mu_S$ are the poloidal and toroidal potentials of
the source $\w{S(\beta)}$ [they can thus be determined from 
$\w{S(\beta)}$ by formulas 
(\ref{e:lapang_eta_bet})-(\ref{e:lapang_mu_bet}) with $\beta^{\hat i}$
replaced by $S(\w{\beta})^{\hat i}$].

Having reduced the complicated coupled PDE system 
(\ref{e:poisson_beta_r})-(\ref{e:poisson_beta_p}) to Poisson type equations
(\ref{e:Poisson_Theta_bet}), (\ref{e:beta_r_4}), (\ref{e:Poisson_chi_bet}),
(\ref{e:Poisson_eta_bet}) and (\ref{e:Poisson_mu_bet}), 
various strategies can be devised to get the solution. 
In all of them, we take advantage of the fact that the Poisson equation 
(\ref{e:Poisson_mu_bet}) for the toroidal part is fully decoupled from 
the others to solve it first and hence get $\mu$.
Similarly the Poisson equation 
(\ref{e:Poisson_Theta_bet}) for the divergence is decoupled from the 
other equations. So we can solve it to get $\Theta$. 
Then we plug $\Theta$ on the right-hand side of Eq.~(\ref{e:beta_r_4}) 
and solve it to get $\beta^r$. An alternative approach 
is to solve the
Poisson equation (\ref{e:Poisson_chi_bet}) for $\chi$ and obtain
$\beta^r$ as $\chi/r$.
Then we have the following options: (i) deduce 
$\eta$ from Eq.~(\ref{e:lapang_eta_bet}); (ii) solve the Poisson 
equation (\ref{e:Poisson_eta_bet}) to get $\eta$. 
Method (ii) requires to solve an additional Poisson equation, while 
method (i) requires only a division by $-\ell(\ell+1)$ of the coefficients of 
spherical harmonics expansions, making a total of three scalar Poisson 
equations to solve the system. 
However method (i) involves the radial 
derivative of $\beta^r$ which may result in a low order of differentiability
of the numerical solution. 

\subsection{Resolution of the tensor wave equation} 
\label{s:resol_tensor_wave}

\subsubsection{Spherical components}

By means of the derivation formula (\ref{e:cDT_ortho}) 
with the explicit values (\ref{e:Ricci_rot_spher}) of the connection
coefficients, the tensor wave equation~(\ref{e:dalembert_hij_TT})
can be written explicitly in terms of the components 
${\bar h}^{\hat i\hat j}$ of the TT part of $\w{h}$ 
with respect to the orthonormal spherical
basis:
\begin{widetext}
\bea
  - \dder{{\bar h}^{rr}}{t}  & + & \dder{{\bar h}^{rr}}{r} + 
  {2\over r} \der{{\bar h}^{rr}}{r}
  +{1\over r^2} \Bigg[ \Delta_{\th\ph} {\bar h}^{rr}
  - 4 {\bar h}^{rr}  - 4 \der{{\bar h}^{r\th}}{\th} - {4 {\bar h}^{r\th}\over\tan\th} 
  - {4\over\sin\th} \der{{\bar h}^{r\ph}}{\ph}  + 2 {\bar h}^{\th\th} + 2 {\bar h}^{\ph\ph} 
  \Bigg]  = {\bar S}^{rr} , \label{e:dalemb_hrr} \\
  - \dder{{\bar h}^{r\th}}{t} & + &  \dder{{\bar h}^{r\th}}{r} + {2\over r} \der{{\bar h}^{r\th}}{r}
  +{1\over r^2}\Bigg[ \Delta_{\th\ph} {\bar h}^{r\th} 
  - \left( 4+{1\over\sin^2\th}	\right) {\bar h}^{r\th}  
  +2\der{{\bar h}^{rr}}{\th} - 2\der{{\bar h}^{\th\th}}{\th}
  -2{\cos\th\over\sin^2\th} \der{{\bar h}^{r\ph}}{\ph} 
  -{2{\bar h}^{\th\th}\over\tan\th} \nonumber \\
  & &- {2\over\sin\th}\der{{\bar h}^{\th\ph}}{\ph}
  + {2 {\bar h}^{\ph\ph}\over \tan\th} \Bigg] = {\bar S}^{r\th} , \label{e:dalemb_hrt} \\
  - \dder{{\bar h}^{r\ph}}{t} & + &  \dder{{\bar h}^{r\ph}}{r} + {2\over r} \der{{\bar h}^{r\ph}}{r}
  +{1\over r^2}\Bigg[ \Delta_{\th\ph} {\bar h}^{r\ph}
  - \left( 5+{1\over\tan^2\th}	\right) {\bar h}^{r\ph} 
   + {2\over\sin\th} \der{{\bar h}^{rr}}{\ph}
  + 2 {\cos\th\over\sin^2\th} \der{{\bar h}^{r\th}}{\ph}
  - 2 \der{{\bar h}^{\th\ph}}{\th} \nonumber \\
  & &- {2\over\sin\th} \der{{\bar h}^{\ph\ph}}{\ph}
  - {4{\bar h}^{\th\ph}\over\tan\th} \Bigg] = {\bar S}^{r\ph}, \label{e:dalemb_hrp}\\
  - \dder{{\bar h}^{\th\th}}{t} & + &  \dder{{\bar h}^{\th\th}}{r} + {2\over r} \der{{\bar h}^{\th\th}}{r}
  +{1\over r^2}\Bigg[ \Delta_{\th\ph} {\bar h}^{\th\th} 
  - {2{\bar h}^{\th\th}\over\sin^2\th}  + 4 \der{{\bar h}^{r\th}}{\th} 
  - 4 {\cos\th\over\sin^2\th} \der{{\bar h}^{\th\ph}}{\ph}
  + 2 {\bar h}^{rr} + {2{\bar h}^{\ph\ph}\over\tan^2\th} \Bigg] = {\bar S}^{\th\th}, 
  		\label{e:dalemb_htt}\\
  - \dder{{\bar h}^{\th\ph}}{t}& + &  \dder{{\bar h}^{\th\ph}}{r} + {2\over r} \der{{\bar h}^{\th\ph}}{r}
  +{1\over r^2}\Bigg[ \Delta_{\th\ph} {\bar h}^{\th\ph}
  - 2 \left( 1+{2\over\tan^2\th} \right) {\bar h}^{\th\ph} 
   + {2\over\sin\th} \der{{\bar h}^{r\th}}{\ph} 
  + 2 \der{{\bar h}^{r\ph}}{\th}\nonumber \\
  & &
  + 2{\cos\th\over\sin^2\th} \left( \der{{\bar h}^{\th\th}}{\ph}
  	-\der{{\bar h}^{\ph\ph}}{\ph} \right) 
	- {2 {\bar h}^{r\ph}\over \tan\th} \Bigg] = {\bar S}^{\th\ph}, 
	\label{e:dalemb_htp} \\
  - \dder{{\bar h}^{\ph\ph}}{t} & + & \dder{{\bar h}^{\ph\ph}}{r} + {2\over r} \der{{\bar h}^{\ph\ph}}{r}
  +{1\over r^2}\Bigg[ \Delta_{\th\ph} {\bar h}^{\ph\ph}
  - {2 {\bar h}^{\ph\ph}\over\sin^2\th} + {4\over\sin\th} \der{{\bar h}^{r\ph}}{\ph} 
  + 4 {\cos\th\over\sin^2\th} \der{{\bar h}^{\th\ph}}{\ph}
  + 2 {\bar h}^{rr} + {2 {\bar h}^{\th\th}\over \tan^2\th} \nonumber \\
  & & 
  + {4 {\bar h}^{r\th}\over\tan\th} \Bigg] = {\bar S}^{\ph\ph}, \label{e:dalemb_hpp}	
\eea
where ${\bar S}^{\hat i\hat j}$ denotes the right-hand side of 
Eq.~(\ref{e:dalembert_hij_TT}) : 
${\bar S}^{\hat i\hat j} := {\bar\sigma}^{\hat i\hat j} 
+ (L\dot\beta)^{\hat i\hat j}$.
These equations must be supplemented by the TT conditions 
[Eq.~(\ref{e:hij_TT})], which read, in term
of components with respect to $(\w{e}_{\hat i})$,
\bea
&& \der{{ \bar h}^{rr}}{r} + {2{ \bar h}^{rr}\over r}
  + {1\over r} \left[ \der{{ \bar h}^{r\th}}{\th} 
  + {1\over \sin\th} \der{{ \bar h}^{r\ph}}{\ph}
  - {\bar h}^{\th\th} - { \bar h}^{\ph\ph} + {{ \bar h}^{r\th}\over \tan\th} \right]  \ = \ 0 
  				\label{e:Dirac_ortho_r}\\ 
&& \der{{ \bar h}^{r\th}}{r} + {3{ \bar h}^{r\th}\over r}
  + {1\over r} \left[ \der{{ \bar h}^{\th\th}}{\th} 
  + {1\over \sin\th} \der{{ \bar h}^{\th\ph}}{\ph}
  + {1 \over \tan\th} \left( { \bar h}^{\th\th} - { \bar h}^{\ph\ph} \right) \right]
    \ = \ 0 \label{e:Dirac_ortho_t} \\
&& \der{{ \bar h}^{r\ph}}{r} + {3{ \bar h}^{r\ph}\over r}
  + {1\over r} \left[  \der{{ \bar h}^{\th\ph}}{\th} 
  + {1\over \sin\th} \der{{ \bar h}^{\ph\ph}}{\ph}
  + {2 {\bar h}^{\th\ph}\over \tan\th} \right] \ = \ 0 , 
                                            \label{e:Dirac_ortho_p} \\
&& {\bar h}^{rr} + {\bar h}^{\th\th} + {\bar h}^{\ph\ph} = 0 .    
                \label{e:traceless_hbar}                                        
\eea
\end{widetext}
As discussed in Sec.~\ref{s:decomp_TT}, the TT conditions and the
$\square$ operator commute, so 
provided that the source $\w{\bar S}$ is TT, the solution 
$\w{\bar h}$ will also be TT.  

For the steady state case ($\dert{}{t}=0$) or for an implicit time 
scheme\footnote{With Chebyshev spectral methods, 
the accumulation of collocation points 
near the boundaries implies a very severe Courant-Friedrich-Levy condition
and in practice prohibits explicit schemes.}, we need to invert the full 
operator on the left hand side of the system
(\ref{e:dalemb_hrr})-(\ref{e:dalemb_hpp}).
One immediately notices that this system couples all the components
$h^{\hat i\hat j}$. 

A natural idea to solve the system (\ref{e:dalemb_hrr})-(\ref{e:dalemb_hpp})
would be to expand $\w{\bar h}$
onto a a basis of {\em tensor} spherical harmonics. 
Notice that, contrarily to {\em scalar} spherical
harmonics, there are several types of tensor ones (for a
review, see \cite{Thorn80}). A first family
has been introduced by Mathews \cite{Mathe62} and Zerilli
\cite{Zeril70}; they are called {\em pure orbital} harmonics in
\cite{Thorn80} and are eigenvectors of the angular Laplacian
(\ref{e:angu_Lap}) acting on tensors. A second family is made of {\em
pure spin} harmonics \cite{ReggeW57,Zeril70} which are very well suited for
describing gravitational radiation in the radiation zone (where one
supposes that the wave vector is parallel to the radial direction).
However, it should be realized that all families of tensor spherical harmonics
are based on a longitudinal/transverse decomposition with a notion 
of {\em transversality} different from the one used here: in our 
acceptation, {\em transverse} means {\em divergence-free} 
[Eqs.~(\ref{e:Dirac_div_h}) and (\ref{e:hij_TT})], whereas in tensor
spherical harmonics literature, {\em transverse} means {\em orthogonal
with respect to the radial vector $\w{e}_r$}. Asymptotically both notions
coincide, but this is not the case at finite $r$. 
From the very definition of Dirac gauge [Eqs.~(\ref{e:Dirac_div_h})], it
is clear that the notion of transversality relevant to our problem is
the divergence-free one. 
As shown by Mathews \cite{Mathe62} and explicited in the quadrupolar case by 
Teukolsky \cite{Teuko82}, it is possible a form linear combinations 
of tensor spherical harmonics which are divergence-free. 
We propose here a different route, which is actually simpler.
We do not perform any expansion
onto the tensor spherical harmonics, but use directly the  
traceless and divergence-free properties to reduce the tensor wave equation
to two scalar wave equations, reflecting the two degree of freedoms of
a TT symmetric tensor. 

Before presenting this method, let us comment upon another tentative of
decoupling the system (\ref{e:dalemb_hrr})-(\ref{e:dalemb_hpp}) that
one might naively contemplate. It would consist in 
solving 
separately each equation (\ref{e:dalemb_hrr}),...,(\ref{e:dalemb_hpp})
by treating as source the terms with 
${\bar h}^{\hat k \hat l}$ ($k\not= i$ or $l\not = j$) so that 
only an operator acting on the component ${\bar h}^{\hat i\hat j}$ 
would appear on the left-hand side. Of course, since 
the other components of 
$\w{\bar h}$ would be present on the right-hand side, such a method would
require some iteration. 
However this method is not applicable, due to 
the bad behavior of the truncated operator
(i.e. the operator which acts only on ${\bar h}^{\hat i \hat j}$ in the
component ${\hat i \hat j}$): for a regular source, it gives a non-regular
solution. Take for instance Eq.~(\ref{e:dalemb_hrr}) in
the stationary case ($\dert{}{t}=0$): the operator acting on
${\bar h}^{rr}$ is 
\be \label{e:operator_hrr}
 {\cal O} {\bar h}^{rr}  := \dder{{\bar h}^{rr}}{r} + 
  {2\over r} \der{{\bar h}^{rr}}{r}
  +{1\over r^2} \left( \Delta_{\th\ph} {\bar h}^{rr}
  - 4 {\bar h}^{rr} \right)  .	
\ee
Now ${\bar h}^{rr} = \chi / r^2$, where 
$\chi = f_{ik} f_{jl} {\bar h}^{ij} r^k r^l $ is a regular scalar field
on $\Sigma_t$ [see Eq.~(\ref{e:def_chi}) below]. 
${\bar h}^{rr}$ is therefore expandable in scalar 
spherical harmonics $Y_\ell^m(\th,\ph)$. For a given $(\ell,m)$, the
behavior of ${\bar h}^{rr}$ near the origin $r=0$ must therefore be
\be
	{\bar h}^{rr} \sim r^n \, Y_\ell^m(\th,\ph) ,
\ee
where $n$ is some positive integer, in order for ${\bar h}^{rr}$ to
be regular. 
Inserting this expression into Eq.~(\ref{e:operator_hrr}) results in 
\be
	{\cal O} {\bar h}^{rr} = \left[ n(n-1) + 2 n - \ell(\ell+1) -4\right] 
	r^{n-2} \, Y_\ell^m(\th,\ph) . 
\ee
Thus we get a regular solution of the homogeneous equation ${\cal O} {\bar h}^{rr}=0$
near $r=0$ only if, for any $\ell$, there exists a 
strictly positive integer $n$ such that 
$n^2 + n - \ell(\ell+1) -4 = 0$. However in general, this 
last equation does not admit any integer solution $n$. 
The generalization to the time-dependent case is straightforward. 
Moreover, even if $r=0$ is excluded from the computational domain
(for example when treating black holes), a similar regularity problem
appears in the other equations on the axis $\th=0$ or $\pi$.

\subsubsection{Reduction to two scalar wave equations}
\label{s:reduc}

As mentioned above, it is possible to use the four TT conditions 
(\ref{e:Dirac_ortho_r})-(\ref{e:traceless_hbar}) to 
decouple the system (\ref{e:dalemb_hrr})-(\ref{e:dalemb_hpp})
and to reduce it to the resolution of two 
scalar wave equations. 

A first way to proceed is to manipulate directly equations
(\ref{e:dalemb_hrr})-(\ref{e:traceless_hbar}). For instance,  
inserting the first divergence-free condition (\ref{e:Dirac_ortho_r}) into
(\ref{e:dalemb_hrr}) and  using the traceless condition 
(\ref{e:traceless_hbar}) results in the disappearing of the terms
involving ${\bar h}^{r\th}$, ${\bar h}^{r\ph}$, ${\bar h}^{\th\th}$ and 
${\bar h}^{\ph\ph}$:
\be
   - \dder{{\bar h}^{rr}}{t} + \dder{{\bar h}^{rr}}{r} + 
  {6\over r} \der{{\bar h}^{rr}}{r}
  +{1\over r^2} \left( \Delta_{\th\ph} {\bar h}^{rr}
  + 6 {\bar h}^{rr} \right)  = {\bar S}^{rr} .  \label{e:evol_hrr}
\ee

To perform a more systematic treatment, as well as to gain some insight,
it is worth to introduce the scalar product (with respect to $\w{f}$) 
of $\w{\bar h}$ with the position vector $\w{r}$
defined by Eq.~(\ref{e:r_def}):
\be
    V^i := f_{kl} \, {\bar h}^{ik} r^l ,
\ee
or, in term of components, 
\be 
    V^{\hat i} = (r {\bar h}^{rr}, r {\bar h}^{r\th},
        r {\bar h}^{r\ph}) .  \label{e:V_comp}
\ee 
Note that the vector field $\w{V}$ thus defined is regular  
(for $\w{f}$, $\w{\bar h}$ and $\w{r}$ are regular tensor fields on $\Sigma_t$). 
From the identities $\square V^i = f_{kl} r^l \square {\bar h}^{ik} 
+ 2 \cD_k {\bar h}^{ik}$ and $\cD_i V^i = f_{kl} r^l \cD_i {\bar h}^{ik}
+ f_{ij} {\bar h}^{ij}$ and the TT character of $\w{\bar h}$, we deduce 
immediately that the $(rr, r\th, r\ph)$ part of the system 
(\ref{e:dalemb_hrr})-(\ref{e:dalemb_hpp}) with the TT conditions
(\ref{e:Dirac_ortho_r})-(\ref{e:traceless_hbar}) is equivalent to
the vector wave equation
\be
     \square V^i = f_{kl} {\bar S}^{ik} r^l \quad \mbox{with}
     \quad \cD_i V^i = 0 .  \label{e:vector_wave}
\ee
Let us introduce the (regular) scalar 
field $\chi$ \footnote{we use the same notation $\chi$
as for the decomposition of the shift vector in Sec.~\ref{s:vector_shift},
assuming that no confusion may arise.} as the scalar product (with respect
to $\w{f}$) of $\w{r}$ and $\w{V}$, 
\be \label{e:chi_h_def}
    \chi := f_{kl} r^k V^l  = r V^r = r^2 {\bar h}^{rr} . 
\ee
From the identity $\square\chi = f_{kl} r^k \square V^l + 2\cD_k V^k$
and the divergence-free character of $\w{V}$, we see that 
Eq.~(\ref{e:vector_wave}) implies the following scalar wave equation
\be \label{e:dalemb_chi}
	\square \chi = r^2 {\bar S}^{rr}  .
\ee 
Solving this equation immediately provides ${\bar h}^{rr}$ by
\be \label{e:def_chi}
	{\bar h}^{rr} = {\chi \over r^2} .
\ee
Note that inserting this last relation into Eq.~(\ref{e:evol_hrr})
would have lead directly to Eq.~(\ref{e:dalemb_chi}). 

We then proceed as for the vector Poisson equation treated in 
Sec.~\ref{s:vector_shift}, namely we perform the radial/angular
decomposition of $\w{V}$ following
Eq.~(\ref{e:beta_eta_mu})\footnote{again, we use the same notation
$\eta$ and $\mu$ as for the decomposition of $\w{\beta}$
presented in Sec.~\ref{s:vector_shift},
assuming that no confusion may arise.}:
\be
    \w{V} = V^r \w{e}_r + \left[ r \w{\cD} \eta - (\w{e}_r \cdot
    \w{\cD}\eta)\, \w{r} \right] + \w{r} \times \w{\cD}\mu .                      
\ee
Combining the above equation with Eq.~(\ref{e:V_comp}), we see that  
the potentials $\eta$ and $\mu$ are related to the components
${\bar h}^{r\th}$ and ${\bar h}^{\th\th}$ by 
\bea
  {\bar h}^{r\th} & = & {1\over r} \left( \der{\eta}{\th} - {1\over\sin\th}
  	\der{\mu}{\ph} \right) \label{e:def_eta_mu_t} \\
  {\bar h}^{r\ph} & = & {1\over r} \left( {1\over\sin\th} \der{\eta}{\ph}
  + \der{\mu}{\th} \right) . \label{e:def_eta_mu_p}	
\eea
Performing the same decomposition of the source, we get:
\bea
  {\bar S}^{r\th} & = & {1\over r} \left( \der{\eta_{\bar S}}{\th} 
        - {1\over\sin\th} \der{\mu_{\bar S}}{\ph} \right) , \\
  {\bar S}^{r\ph} & = & {1\over r} \left( 
    {1\over\sin\th} \der{\eta_{\bar S}}{\ph}
  + \der{\mu_{\bar S}}{\th} \right) .
\eea
Given ${\bar S}^{r\th}$ and ${\bar S}^{r\ph}$, $\eta_{\bar S}$ and 
$\mu_{\bar S}$ are computed from the analog of 
Eqs.~(\ref{e:lapang_eta_bet})-(\ref{e:lapang_mu_bet}) by 
\bea
 \Delta_{\th\ph} \eta_{\bar S} & = & r\left( \der{{\bar S}^{r\th}}{\th}
 + {{\bar S}^{r\th}\over\tan\th} + {1\over\sin\th} \der{{\bar S}^{r\ph}}{\ph} \right) 
 	\label{e:lapang_Sigma} \\
 \Delta_{\th\ph} \mu_{\bar S} & = & r\left( \der{{\bar S}^{r\ph}}{\th}
 	+ {{\bar S}^{r\ph}\over\tan\th} - {1\over \sin\th} \der{{\bar S}^{r\th}}{\ph}
	\right) \label{e:lapang_calT} .
\eea
As already discussed in Sec.~\ref{s:vector_shift}, the potentials 
$\eta_{\bar S}$ and $\mu_{\bar S}$
are expandable in scalar spherical harmonics $Y_\ell^m(\th,\ph)$.
Equations (\ref{e:lapang_Sigma})-(\ref{e:lapang_calT})
are then algebraic ($\Delta_{\th\ph} u
\rightarrow - \ell(\ell+1) u$) in terms of the coefficients of the
spherical harmonics expansion. 

The angular part of the vector wave equation (\ref{e:vector_wave}) is
equivalent to the following system, analogous to 
Eqs.~(\ref{e:Poisson_eta_bet})-(\ref{e:Poisson_mu_bet}) with $\Theta=0$
(since $\w{V}$ is divergence-free) and $V^r = r {\bar h}^{rr}$:
\bea
	\square\eta & = & \eta_{\bar S} - {2 {\bar h}^{rr}\over r}  , 
                                \label{e:dalemb_eta} \\
	\square\mu & = &\mu_{\bar S} \label{e:dalemb_mu}.
\eea
We can see here that the equation for $\mu$ is fully decoupled
from the other equations, contrarily to that for $\eta$ which contains
${\bar h}^{rr}$. Actually the divergence-free condition $\cD_i V^i=0$
relates $\eta$ to ${\bar h}^{rr}$ by Eq.~(\ref{e:lapang_eta_bet}) 
(with $V^r = r {\bar h}^{rr} = \chi/r$):
\be \label{e:lap_ang_eta}
    \Delta_{\th\ph} \eta = - r \left(  r \der{{\bar h}^{rr}}{r} 
    + 3 {\bar h}^{rr} \right) = - \der{\chi}{r} - {\chi\over r}. 
\ee
This last equation can be used to compute $\eta$, once ${\bar h}^{rr}$
has been obtained as the solution of (\ref{e:evol_hrr}) [or from the 
system (\ref{e:dalemb_chi})-(\ref{e:def_chi})], instead of solving the
wave equation (\ref{e:dalemb_eta}). 

At this stage, there remains to compute the angular components
${\bar h}^{\th\th}$, ${\bar h}^{\th\ph}$ and ${\bar h}^{\ph\ph}$.
They can be deduced fully from the other components, by means of
the TT relations (\ref{e:Dirac_ortho_t})-(\ref{e:traceless_hbar}).
Indeed, using the traceless condition (\ref{e:traceless_hbar}), 
the transverse conditions (\ref{e:Dirac_ortho_t}) and (\ref{e:Dirac_ortho_p})
can be written as 
 \bea
   \der{}{\th} (\sin^2\th \: {\bar h}^{\ph\ph}) 
   - {1\over\sin\th}\der{}{\ph} (\sin^2\th\:  {\bar h}^{\th\ph}) & = & T^\th , 
   		\label{e:dF/dth} \\
  {1\over\sin\th} \der{}{\ph} (\sin^2\th \: {\bar h}^{\ph\ph})
  + \der{}{\th} (\sin^2\th\:  {\bar h}^{\th\ph})
        & =& T^\ph , \label{e:dF/dph}
\eea
with 
\bea
   T^\th & := &
   \sin^2\th \left( r\der{{\bar h}^{r\th}}{r} + 3 {\bar h}^{r\th}
   - \der{{\bar h}^{rr}}{\th} - {{\bar h}^{rr}\over\tan\th} \right) , 
                        \label{e:def_T_th}\\
  T^\ph & := &
  - \sin^2\th \left( r\der{{\bar h}^{r\ph}}{r} + 3 {\bar h}^{r\ph} \right) .
                                    \label{e:def_T_ph}
\eea
Taking the angular divergence and the angular curl of 
Eqs.~(\ref{e:dF/dth})-(\ref{e:dF/dph}), as in 
Eqs.~(\ref{e:lapang_Sigma})-(\ref{e:lapang_calT}), we get the system
\bea
	\Delta_{\th\ph} (\sin^2\th \: {\bar h}^{\ph\ph}) & = & \der{T^\th}{\th}
 + {T^\th\over\tan\th} + {1\over\sin\th} \der{T^\ph}{\ph} \label{e:lap_ang_F} \\
 \Delta_{\th\ph} (\sin^2\th\:  {\bar h}^{\th\ph}) & = &  \der{T^\ph}{\th} + {T^\ph\over\tan\th} 
                - {1\over \sin\th} \der{T^\th}{\ph} \label{e:lap_ang_G} . 
\eea
Again, this system is algebraic in the spherical harmonics representation,
and therefore can be easily solved to get 
$\sin^2\th \: {\bar h}^{\ph\ph}$ and 
$\sin^2\th\:  {\bar h}^{\th\ph}$, after 
$T^\th$ and $T^\ph$ have been evaluated by means of 
Eqs.~(\ref{e:def_T_th})-(\ref{e:def_T_ph}).
The components ${\bar h}^{\ph\ph}$ and ${\bar h}^{\th\ph}$
are then obtained by a division by $\sin^2\th$. 
Finally ${\bar h}^{\th\th}$ is obtained
by the traceless condition (\ref{e:traceless_hbar}): 
\be \label{e:traceless_hat_h}
	{\bar h}^{\th\th} = - {\bar h}^{\ph\ph} - {\bar h}^{rr} . 
\ee

In conclusion we propose to solve the tensor wave equation
(\ref{e:dalembert_hij_TT}) by solving two scalar wave equations:
for $\chi$ [Eq.~(\ref{e:dalemb_chi})] and for $\mu$ 
[Eq.~(\ref{e:dalemb_mu})]. ${\bar h}^{rr}$ is then obtained by dividing 
$\chi$ by $r^2$ [Eq.~(\ref{e:def_chi})]. $\eta$ is obtained from $\chi$
by the quasi-algebraic equation (\ref{e:lap_ang_eta}). From $\mu$ and $\eta$,
we compute ${\bar h}^{r\th}$ and ${\bar h}^{r\ph}$ from 
Eqs.~(\ref{e:def_eta_mu_t})-(\ref{e:def_eta_mu_p}). Then solving the
quasi-algebraic equations (\ref{e:lap_ang_F}) and (\ref{e:lap_ang_G})
gives ${\bar h}^{\ph\ph}$ and ${\bar h}^{\th\ph}$. Finally 
${\bar h}^{\th\th}$ is computed
by the traceless condition (\ref{e:traceless_hat_h}).
The advantage of this procedure consists in solving only for two scalar
wave equations which are linearly decoupled. This guarantees numerical
stability, at least in the linear case.  

\subsubsection{Asymptotic behavior}

Providing that the source ${\bar S}^{ij}$ is decaying sufficiently fast, 
the general asymptotic outgoing solutions of the two scalar wave equations
to be solved, Eqs.~(\ref{e:dalemb_chi}) and (\ref{e:dalemb_mu}), have the form 
\be
\chi \sim {1\over r} {\cal F}_\chi(t - r,\th,\ph) \quad\mbox{and}\quad
\mu \sim {1\over r} {\cal F}_\mu(t - r,\th,\ph) ,
\ee
where ${\cal F}_\chi$ and ${\cal F}_\mu$ are two bounded functions. 
From Eq.~(\ref{e:lap_ang_eta}), we then get the following asymptotic behavior
for the potential $\eta$:
\be
    \eta \sim {1\over r} {\cal F}_\eta(t - r,\th,\ph) ,
\ee
where ${\cal F}_\eta$ is a bounded function. 
The asymptotic behavior of the components ${\bar h}^{rr}$,
${\bar h}^{r\th}$ and ${\bar h}^{r\ph}$ follow immediately from 
Eqs.~(\ref{e:def_chi}), (\ref{e:def_eta_mu_t}) and (\ref{e:def_eta_mu_p}):
\bea
  {\bar h}^{rr} & \sim &  {1\over r^3} {\cal F}_\chi(t - r,\th,\ph) , 
            \label{e:asymptot_hrr} \\
  {\bar h}^{r\th} & \sim &  {1\over r^2} {\cal F}_1(t - r,\th,\ph) , 
            \\
  {\bar h}^{r\ph} & \sim &  {1\over r^2} {\cal F}_2(t - r,\th,\ph) , 
            \label{e:asymptot_hrp}
\eea
where ${\cal F}_1$ and ${\cal F}_2$ are two bounded functions. 
This faster than $O(1/r)$ decay shows 
that the $({\bar h}^{rr},{\bar h}^{r\th}, {\bar h}^{r\ph})$ 
part of $\w{\bar h}$ does not transport any wave, as expected
(cf. the asymptotic TT structure of Dirac gauge discussed in 
Sec.~\ref{s:Dirac_def}).

Thanks to the terms $r\dert{{\bar h}^{r\th}}{r}$ and 
$r\dert{{\bar h}^{r\ph}}{r}$ 
in Eqs.~(\ref{e:def_T_th})-(\ref{e:def_T_ph}), it can be shown 
easily that the asymptotic behavior of ${\bar h}^{\th\ph}$ and 
${\bar h}^{\ph\ph}$ deduced from 
Eqs.~(\ref{e:asymptot_hrr})-(\ref{e:asymptot_hrp}) are
\be
    {\bar h}^{\ph\ph} \sim - {1\over r} h_+(t - r,\th,\ph)
    \quad\mbox{and}\quad 
    {\bar h}^{\th\ph} \sim  {1\over r} h_\times(t - r,\th,\ph) ,
        \label{e:asymptot_hpp}
\ee
where $h_+$ and $h_\times$ are two bounded functions. 
From Eqs.~(\ref{e:traceless_hat_h}), (\ref{e:asymptot_hrr}) and
(\ref{e:asymptot_hpp}), one gets
\be
    {\bar h}^{\th\th} \sim  {1\over r} h_+(t - r,\th,\ph) .
            \label{e:asymptot_htt}
\ee
Contemplating Eqs.~(\ref{e:asymptot_hpp}) and (\ref{e:asymptot_htt}),
we recover the usual behavior of a radiating metric in the TT gauge,
$h_+$ and $h_\times$ being the two gravitational wave modes.

\subsection{Computing the trace $h$ by enforcing the
unit value of the determinant of $\tgm^{\hat i\hat j}$}
\label{s:unit_determ}

Having solved the TT wave equation for $\w{\bar h}$, there remains to 
determine the trace $h=f_{ij} h^{ij}$ 
to reconstruct $\w{h}$ by Eq.~(\ref{e:h_decompTT}),
and then the conformal metric $\w{\tgm} = \w{f} + \w{h}$.
$h$ can be obtained by solving the scalar wave equation (\ref{e:dalembert_h}).
However, $h$ can also be computed in order to enforce a relation
arising from the very definition  of the conformal metric, namely 
that the determinant of the components $\tgm^{ij}$ is equal 
to the inverse of that 
of the flat metric: $\det \tgm^{ij} = f^{-1}$ [cf. Eq.~(\ref{e:dettgm_f})].
It is easy to show this is equivalent to the following requirement 
about the orthonormal components: 
\be \label{e:dettgm_ortho}
	\det \tgm^{\hat i\hat j} = 1 .  
\ee
Replacing $\tgm^{\hat i\hat j}$ by $f^{\hat i\hat j} + h^{\hat i\hat j}$,
this relation writes
\be
	\left| \begin{array}{lll}
		1+ h^{rr} & h^{r\th} & h^{r\ph} \\
		h^{r\th} & 1+ h^{\th\th} & h^{\th\ph} \\
		h^{r\ph} & h^{\th\ph} & 1+ h^{\ph\ph}
		\end{array} \right|  = 1 .
\ee
Expanding the determinant and using 
$h = h^{rr} + h^{\th\th} + h^{\ph\ph}$ results in 
\bea
	 h & = & - h^{rr} h^{\th\th}
	- h^{rr} h^{\ph\ph} - h^{\th\th} h^{\ph\ph}
	+ (h^{r\th})^2 + (h^{r\ph})^2 \nonumber \\*
        &+ &(h^{\th\ph})^2 
        - h^{rr} h^{\th\th} h^{\ph\ph} 
	- 2 h^{r\th} h^{r\ph} h^{\th\ph} + h^{rr} (h^{\th\ph})^2 \nonumber \\*
	& + & h^{\th\th} (h^{r\ph})^2 
        + h^{\ph\ph} (h^{r\th})^2 .
	\label{e:det_unit_ortho1}
\eea
This relation 
shows clearly that among the six components 
$h^{\hat i\hat j}$ only five of them are independent.
The Dirac gauge adds three relations 
between the $h^{\hat i\hat j}$, leaving two independent components:
the two dynamical degrees of freedom of the gravitational field.
Equation (\ref{e:det_unit_ortho1}) shows also that, at the linear order in
$h^{\hat i\hat j}$, the condition $\det\tgm^{\hat i\hat j}=1$ is
equivalent to $h=0$.

We propose to use Eq.~(\ref{e:det_unit_ortho1}) in a numerical code to 
compute $h$, in order to enforce the condition (\ref{e:dettgm_ortho})
by means of the following iterative procedure: initialize 
$h^{\hat i\hat j}$ by the TT part ${\bar h}^{\hat i\hat j}$ 
obtained as a solution of the wave equation (\ref{e:dalembert_hij_TT});
then (i) compute $h$ from Eq.~(\ref{e:det_unit_ortho1}); (ii)
solve the Poisson equation (\ref{e:Lap_Phi_h}) to get $\phi$; 
(iii) insert the values of $h$ and $\phi$ into Eq.~(\ref{e:h_decompTT})
to get $h^{\hat i\hat j}$; (iv) go to (i). 
In practice, this procedure converges up to 
machine accuracy (sixteen digits) within 
at a few iterations. 

\subsection{A constrained scheme for Einstein equations} \label{s:scheme}

Let us sketch the constrained scheme we propose to solve the full 3-D 
time dependent Einstein equations. Our aim here is not to provide
a detailed numerical algorithm, but to show how the Dirac gauge
condition, in conjunction with the use of spherical coordinates,
leads to a method of resolution in which the constraints are automatically
satisfied and the time evolution equations are reduced to only
two scalar wave equations.

At a given time step, one has to solve the two scalar Poisson
equations (\ref{e:lapQ4}) and (\ref{e:evol_K4tr}) to get respectively
$Q$ and $N$, and therefore the conformal factor $\Psi = (Q / N)^{1/2}$.
The Hamiltonian constraint is then automatically satisfied. 
We have outlined the resolution technique of these two scalar Poisson
in Sec.~\ref{s:scalar_poisson}. Let us stress here that a 
very efficient numerical technique to solve within spherical coordinates
scalar Poisson equations with non-compact support has been
presented in Ref.~\cite{GrandBGM01}. 

Then one has to  solve the vector elliptic equation (\ref{e:Poisson_beta})
to get the shift vector $\w{\beta}$, following the procedure 
presented in Sec.~\ref{s:vector_shift}. The momentum constraint is then
automatically satisfied. 

The next equation to be solved is the TT tensor wave equation 
(\ref{e:dalembert_hij_TT}) for $\w{\bar h}$, which arises from the
Einstein dynamical equation (\ref{e:evol_K4st}). As detailed in 
Sec.~\ref{s:resol_tensor_wave}, by fully exploiting the TT character
of $\w{\bar h}$,  
the resolution of this equation is reduced 
to the resolution of two scalar wave equations 
for two scalar potentials $\chi$ and $\mu$
[Eqs.~(\ref{e:dalemb_chi}) and (\ref{e:dalemb_mu})].
From $\chi$ and $\mu$ one can reconstruct all the components 
of $\w{\bar h}$ by taking some derivatives or inverting some
angular Laplacian (which reduces to a mere division by 
$-\ell(\ell+1)$ on spherical harmonics expansions). 

Then the trace $h$ of $\w{h}$ is determined algebraically through 
Eq.~(\ref{e:det_unit_ortho1}) which ensures that 
$\det\tgm_{ij} = f$ [Eq.~[\ref{e:dettgm_f})]. 
From $h$ and $\w{\bar h}$, one reconstructs
$\w{h}$ via Eq.~(\ref{e:h_decompTT}), at the price of solving the
Poisson equation (\ref{e:Lap_Phi_h}) for $\phi$.

Finally, from $\w{h}$, $\w{\beta}$ and $N$, one has to compute 
the conformal extrinsic curvature $A^{ij}$
via Eq.~(\ref{e:Aij_calcul}). 

In the above scheme, the only equations which are not satisfied 
by construction
are (i) Eq.~(\ref{e:kin4_tr}) which relates the time derivative of the
conformal factor $\Psi$ to the divergence of the shift vector $\w{\beta}$
and (ii) Eq.~(\ref{e:dalembert_h}) which is the trace part 
of the wave equation  for $\w{h}$. 
These two scalar equations must however be fulfilled by the solution 
and could be used as evaluators of  the numerical error. 
Alternatively, Eq.~(\ref{e:kin4_tr}) could be enforced as a condition on
$\cD_k \beta^k$ in the resolution of the
elliptic equation (\ref{e:Poisson_beta}) for $\w{\beta}$.

In the above discussion, we have not mentioned the inner boundary conditions
to set on some excised black hole. This point is discussed briefly 
in Appendix~\ref{s:excision} and will be the main subject of a future
study.

\begin{figure*}[htb]
\begin{tabular}{ccc}
\includegraphics[width=0.3\textwidth]{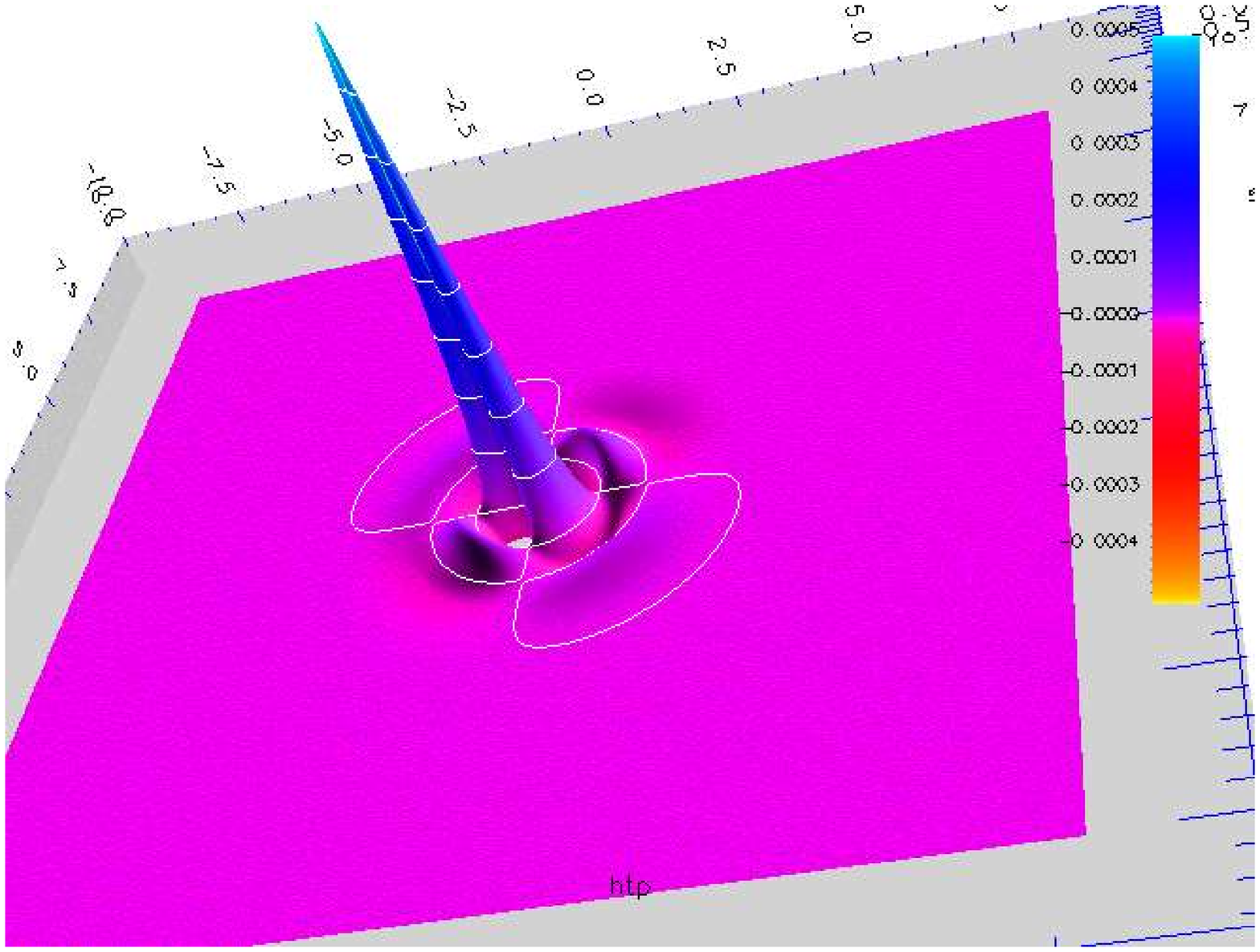} &
\includegraphics[width=0.3\textwidth]{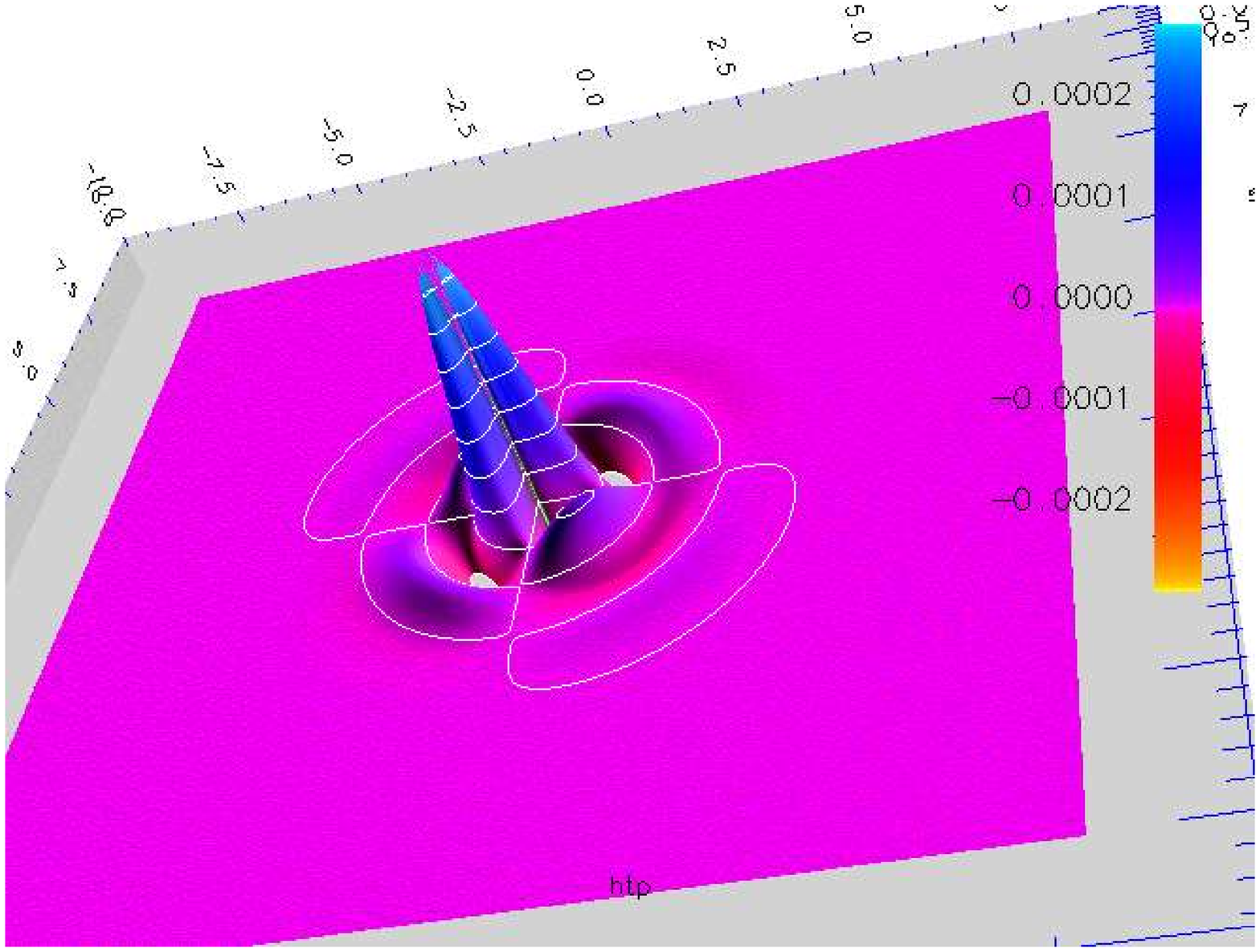} &
\includegraphics[width=0.3\textwidth]{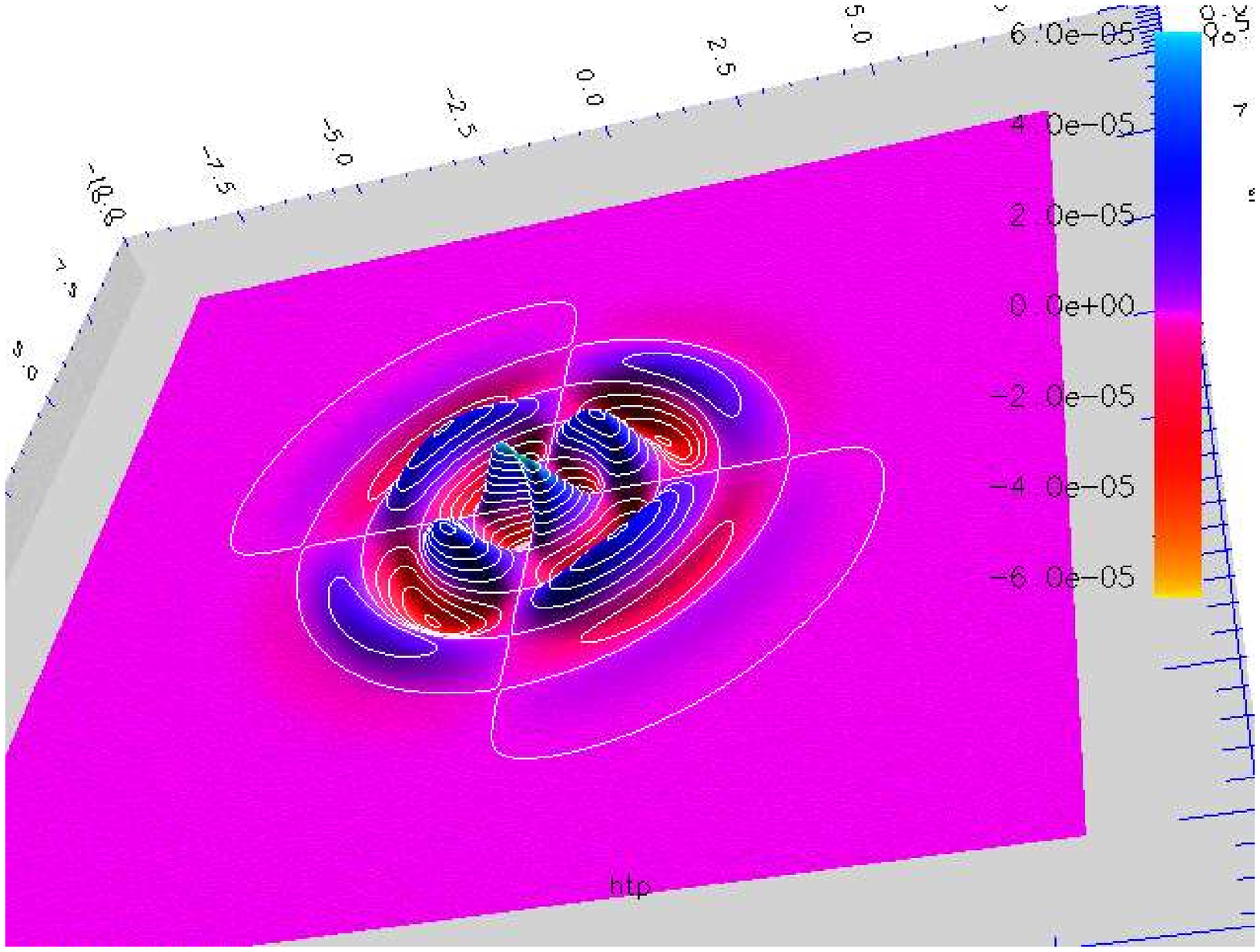} \\
\includegraphics[width=0.3\textwidth]{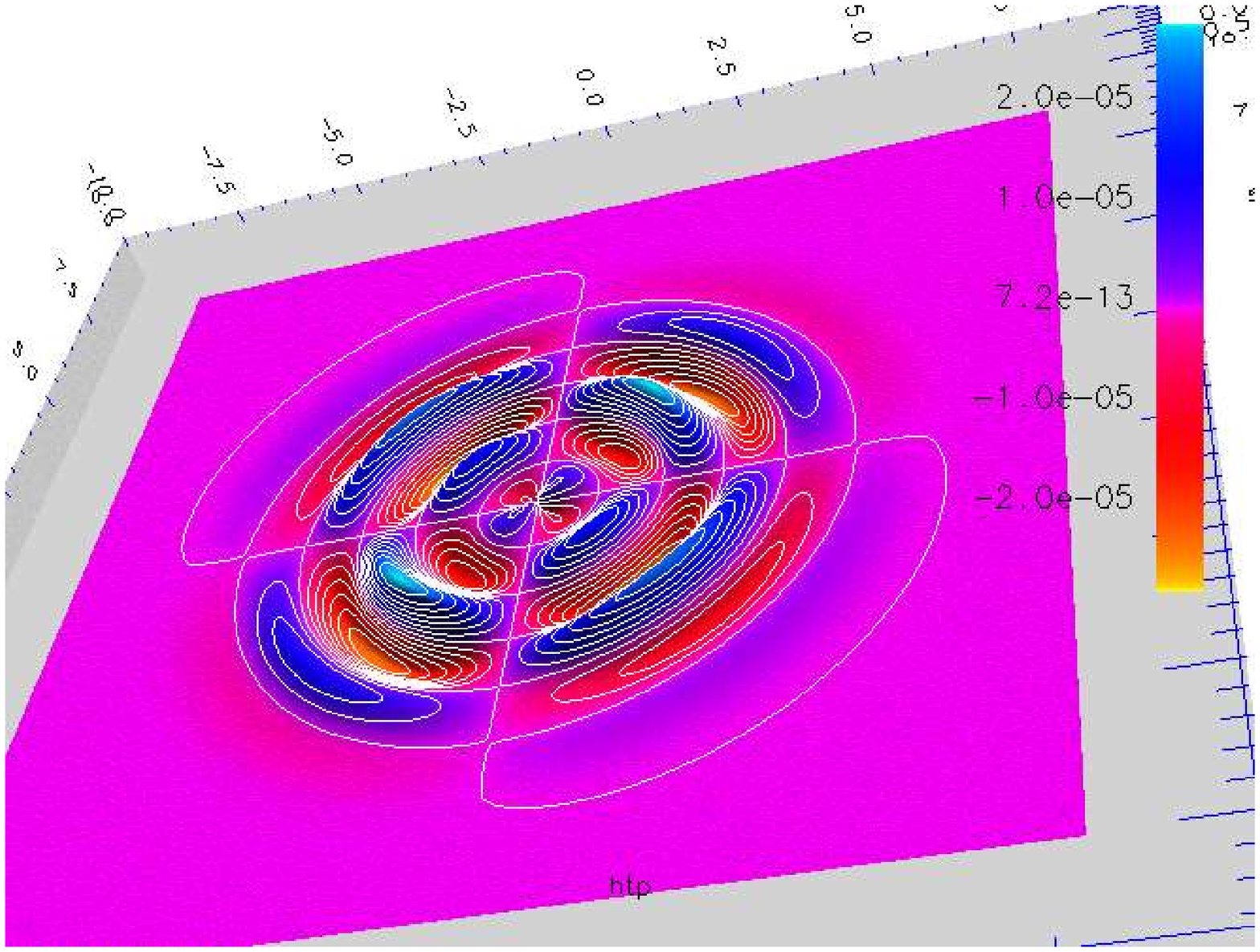} &
\includegraphics[width=0.3\textwidth]{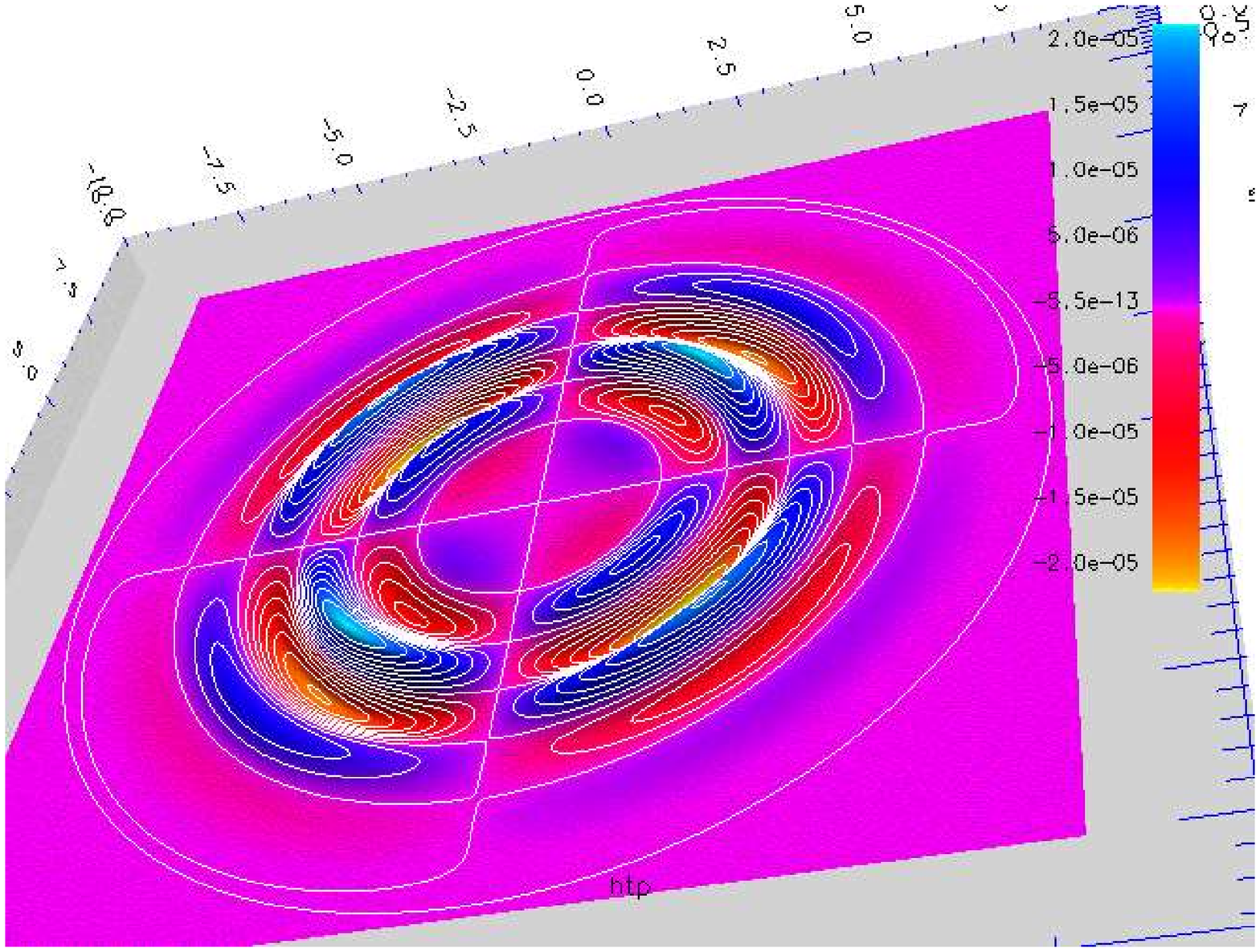} &
\includegraphics[width=0.3\textwidth]{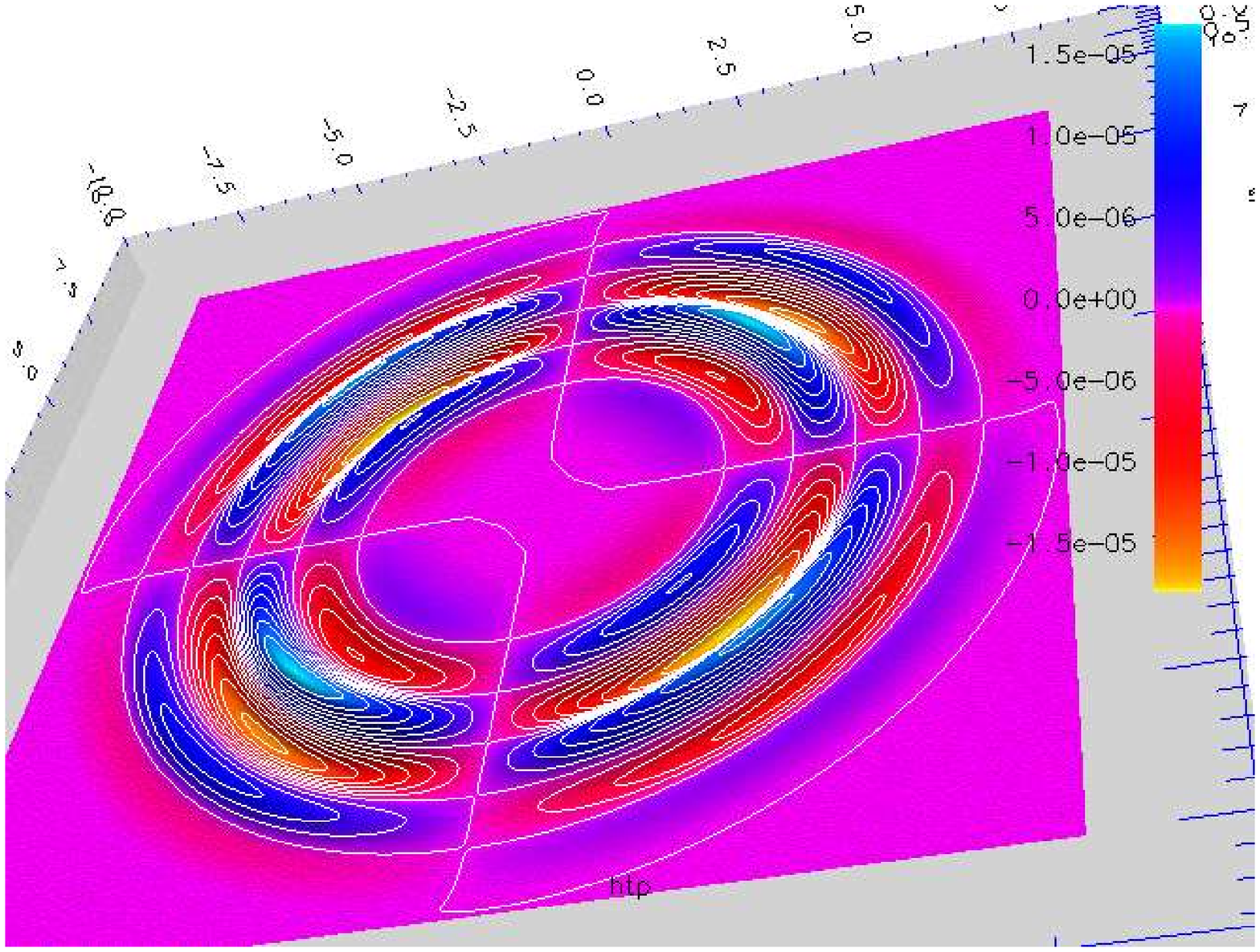} \\
\includegraphics[width=0.3\textwidth]{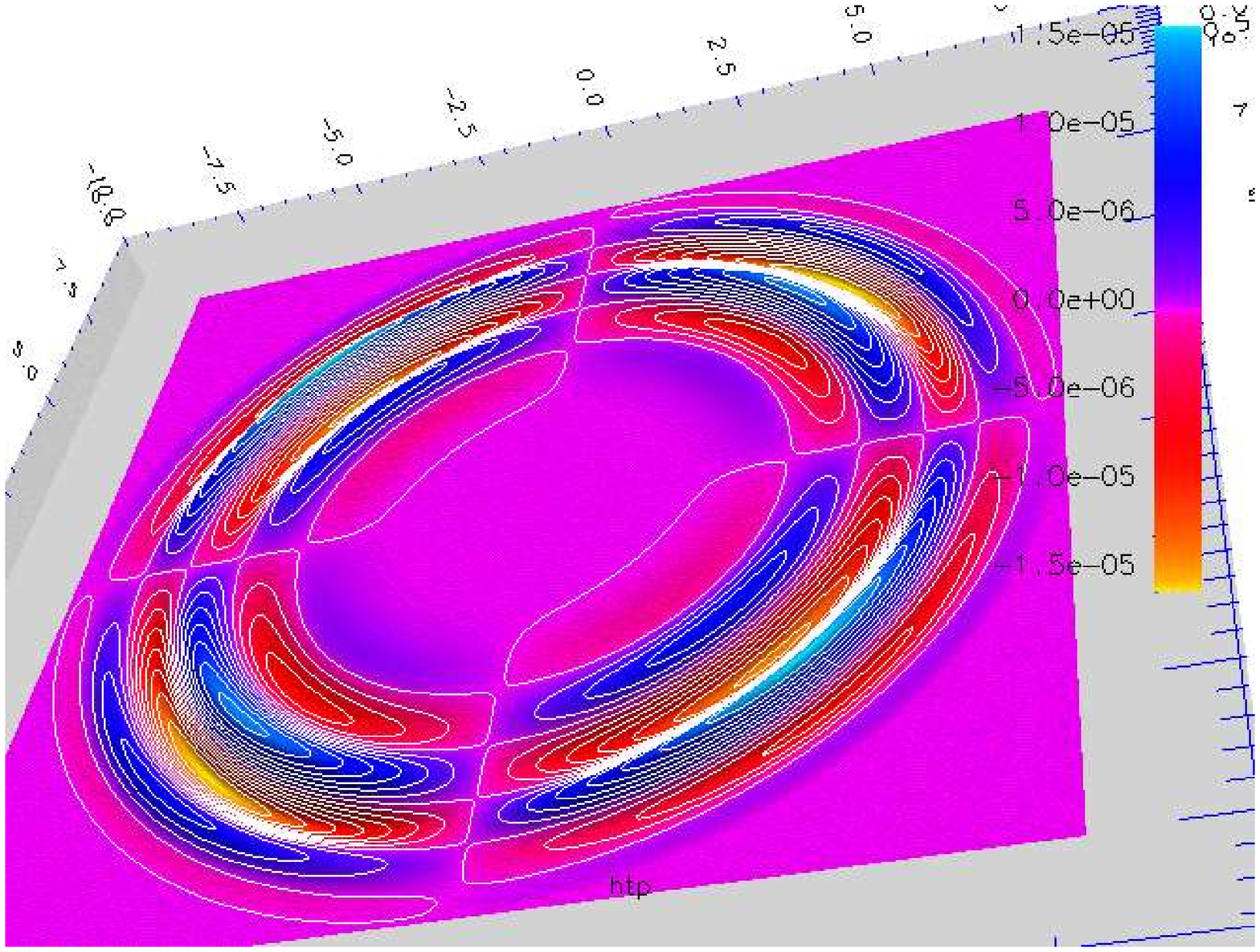} &
\includegraphics[width=0.3\textwidth]{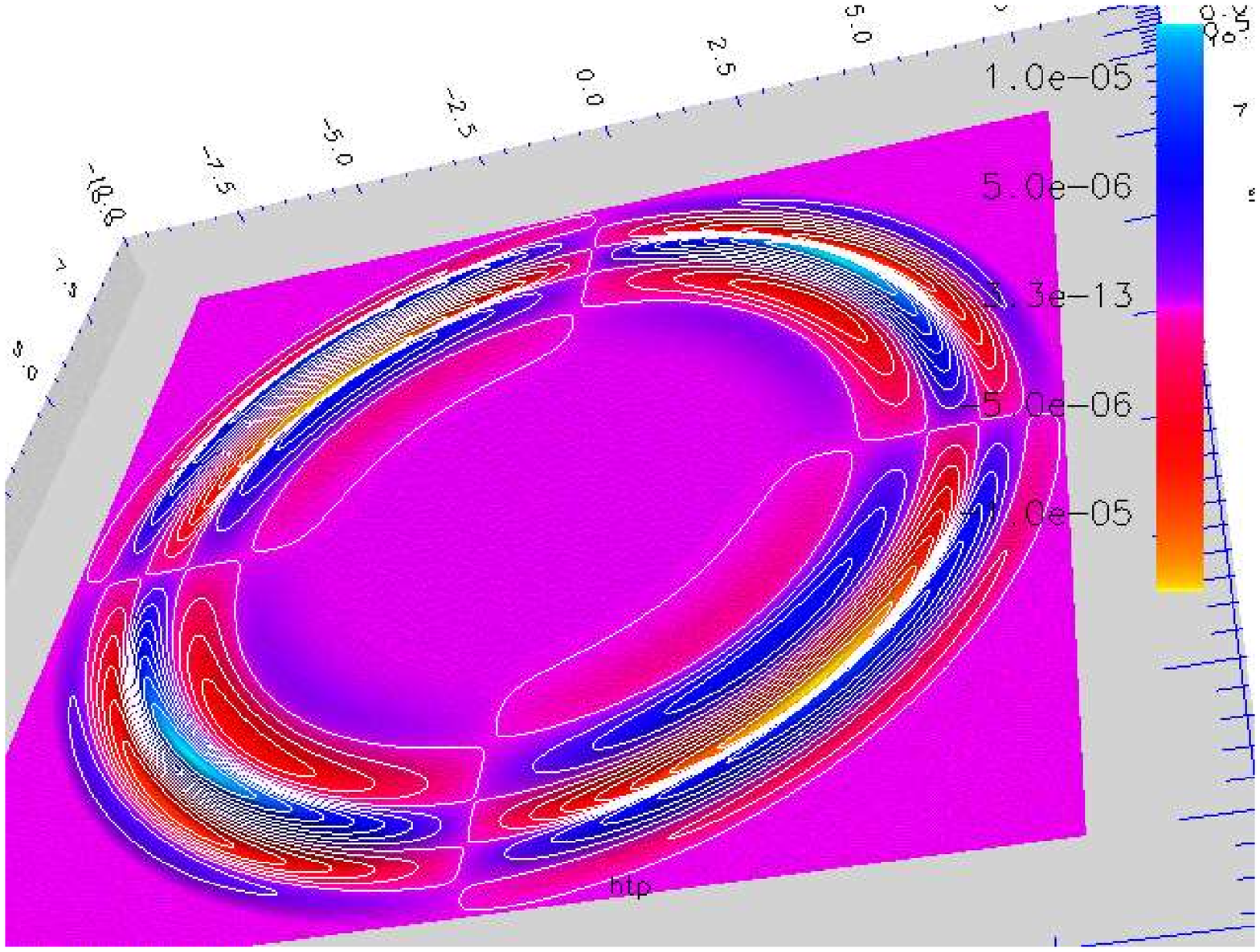} &
\includegraphics[width=0.3\textwidth]{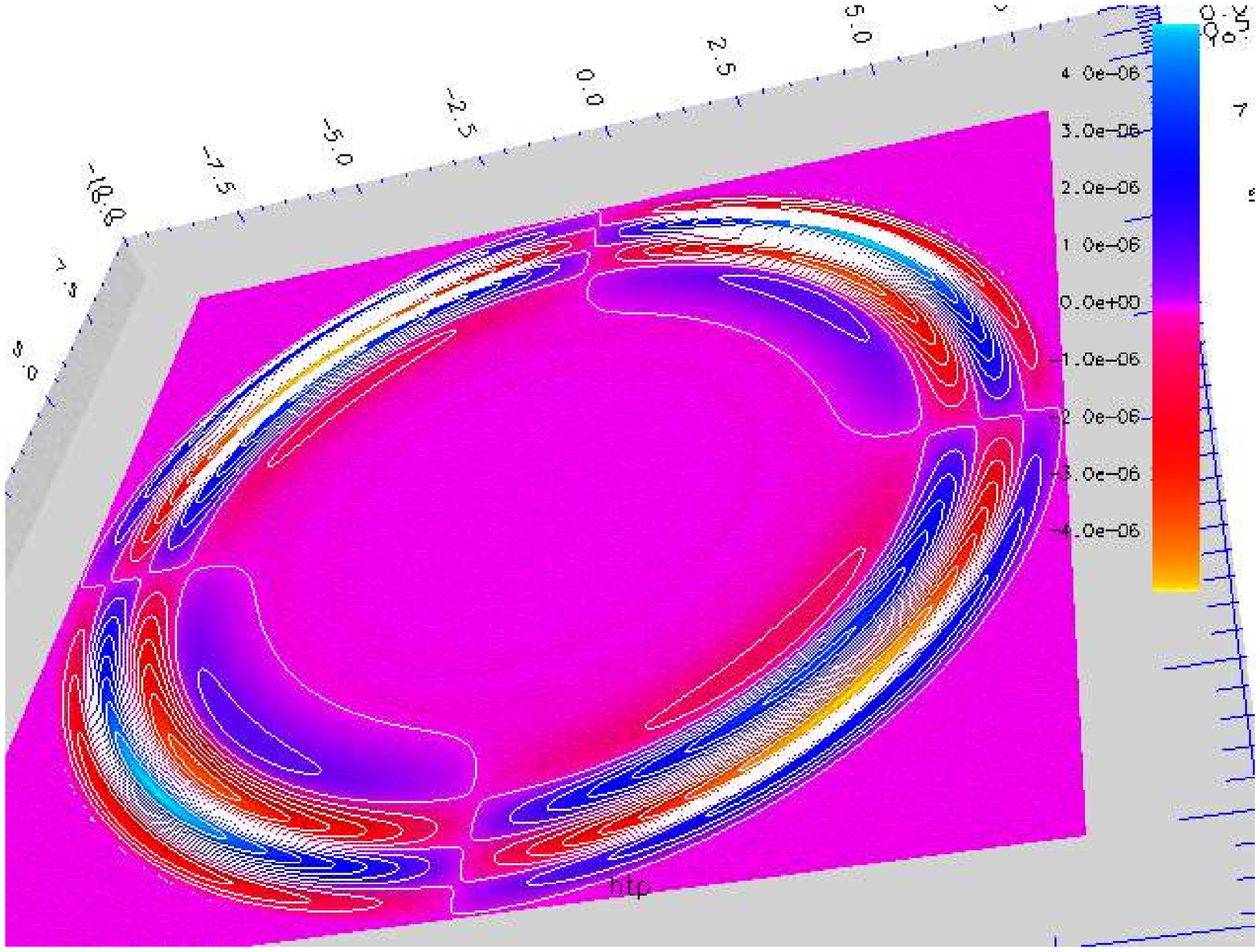} 
\end{tabular}
\caption{ \label{f:hpp}
Evolution of the $h^{\ph\ph}$ component of $\w{h}$ in the plane
$\th=\pi/2$, between $t=0$ (upper left) and $t=8r_0$ (lower right). The various
snapshots are separated by a constant time interval $\Delta t = r_0$.
The size of the depicted square is $16 r_0$, so that the wave extraction surface
at $R_{\rm ext} = 8 r_0$ is given by the circle inscribed in this square.}
\end{figure*}

\section{First results from a numerical implementation}
\label{s:num}

\subsection{Short description of the code}

We have implemented the constrained scheme given in Sec.~\ref{s:scheme}
in a numerical code designed to evolve vacuum spacetimes within 
maximal slicing and Dirac gauge. The code is constructed upon the C++ library
{\sc Lorene} \cite{Lorene}. It uses multidomain spectral methods 
\cite{BonazGM98,BonazGM99b} to 
solve the partial-differential equations within spherical coordinates. 
The scalar Poisson solver is that of Ref.~\cite{GrandBGM01}, whereas the
vector Poisson equation for the shift is solved via the method (ii) 
presented in Sec.~\ref{s:vector_shift}. The scalar wave equations for 
$\chi$ and $\mu$ [Eqs.~(\ref{e:dalemb_chi}) and (\ref{e:dalemb_mu})]
are integrated forward in time by means of the technique presented in 
Ref.~\cite{NovakB04}, namely 
a second-order semi-implicit 
Crank-Nicholson scheme with efficient outgoing-wave boundary conditions.
By ``efficient'' we mean that all
wave modes with spherical harmonics indices $\ell=0$, $1$ and $2$ are
extracted at the outer boundary without any spurious reflection. This is
far better than the Sommerfeld boundary condition commonly used in numerical
relativity and which is valid only for the mode $\ell=0$. 

Various concentric shell-like domains are used, the outermost one being
compactified, to bring spatial infinity to the computational domain. 
The compactified domain is employed to solve all the elliptic equations,
allowing for the correct asymptotic flatness boundary conditions. 
On the contrary, the wave equations are solved only in the non-compactified
domains, the outgoing-wave boundary condition \cite{NovakB04} being imposed at the 
boundary between the last non-compactified shell and the compactified one. 
Further details upon the numerical code will be presented in a future
publication. 

\subsection{Initial data and computational setting}

We have employed the code to evolve pure 3-D gravitational wave spacetimes, 
as in the two BSSN articles \cite{ShibaN95,BaumgS99}. 
Initial data have been obtained by means of the {\em conformal thin sandwich}
formalism \cite{York99,PfeifY03}. 
The freely specifiable parameters of this formalism are $\w{\tgm}$,
$\dert{\w{\tgm}}{t}$, $K$ and $\dert{K}{t}$. In accordance with our choice of
maximal slicing, we set $K=0$ and $\dert{K}{t}=0$. Moreover, 
we use momentarily static data, $\dert{\w{\tgm}}{t}=0$, along with a conformal metric 
$\w{\tgm}$ resulting from  
\bea
    \chi(t=0) &=& {\chi_0\over 2} \, r^2\exp\left(-{r^2\over r_0^2}\right)
                 \, \sin^2\th\, \sin2\ph  \label{e:chi_init} \\
    \mu(t=0) &=& 0 . \label{e:mu_init}
\eea
The constant numbers $\chi_0$ and $r_0$ parametrize respectively the
amplitude and the width of the initial wave packet. 
Let us recall that, within Dirac gauge, the two scalars $\chi$ and $\mu$ 
fully specify $\w{h}$ and thus $\w{\tgm}$: $(\chi,\mu)$ determine a unique
TT tensor $\w{\bar h}$ according to the decomposition presented 
in Sec.~\ref{s:reduc} and the full $\w{h}$ is reconstructed from the trace
$h$ computed in order to ensure $\det\tgm^{ij} = f^{-1}$, following the
method given in Sec.~\ref{s:unit_determ}.
It can be shown that the metric defined by 
Eq.~(\ref{e:chi_init})-(\ref{e:mu_init}) corresponds to an even-parity
Teukolsky wave \cite{Teuko82} with $M=2$.
These initial data are similar to those used by Baumgarte and Shapiro
\cite{BaumgS99} except theirs correspond to a $M=0$ (axisymmetric) 
Teukolsky wave. In particular, we choose an amplitude 
$\chi_0=10^{-3}$ similar to that in Ref.~\cite{BaumgS99}.

A total of 6 numerical domains have been used: a spherical nucleus
of radius $r=r_0$, surrounded by 4 spherical shells of outer radius
$r=2r_0$, $4r_0$, $6r_0$ and $8r_0$, 
and an external compactified domain of inner radius $r=8r_0$.
The outgoing wave boundary conditions discussed above are set at $r=8r_0$,
which we call the {\em wave extraction radius} $R_{\rm ext}$. 
In particular, this means that we do not solve for $\w{h}$ for
$r>8r_0$. Consequently we set $\w{h}$ to zero in the region $r>8r_0$. 
More precisely, we perform a smooth matching of the value of $\w{h}$
at $r=6r_0$ to zero at $r=8r_0$. This means that we solving all the Einstein
equations only for $r<6r_0$. For $r\in[6r_0,\infty)$ 
we are solving the Einstein equations only for the lapse $N$, 
the shift vector $\w{\beta}$ and the conformal factor $\Psi$, with 
$\w{h}$ set to zero in the $r>8r_0$ part of their source terms. 
We take into account the symmetries present in the initial data
(\ref{e:chi_init})-(\ref{e:mu_init}): (i) symmetry with respect to the
plane $\th=\pi/2$ and (ii) symmetry with respect to the transformation
$\ph\mapsto\ph+\pi$. Accordingly, the computational 
coordinate $\th$ spans the interval $[0,\pi/2]$ only 
and $\ph$ the interval $[0,\pi)$.
In each domain, the following numbers of collocations points (= numbers
of polynomials in the spectral expansions)
are used: $N_r\times N_\th\times N_\ph = 17\times 9 \times 8$.
The corresponding memory requirement is 260~MB. This modest value
allows the computation to be performed on a laptop. 
We have used two different time steps $\delta t = 10^{-2} r_0$ and
$\delta t = 5\, 10^{-3} r_0$, to investigate the effects of time
discretization.  

\begin{figure}[htb]
\includegraphics[width=0.45\textwidth]{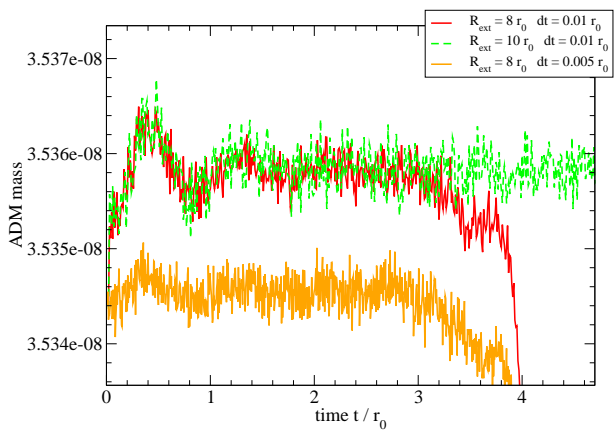} 
\caption{ \label{f:ADM1}
Evolution of the ADM mass for three different computational
settings, corresponding to different values of the time step $\delta t$
and the wave extraction radius $R_{\rm ext}$.}
\end{figure}

\begin{figure}[htb]
\includegraphics[width=0.45\textwidth]{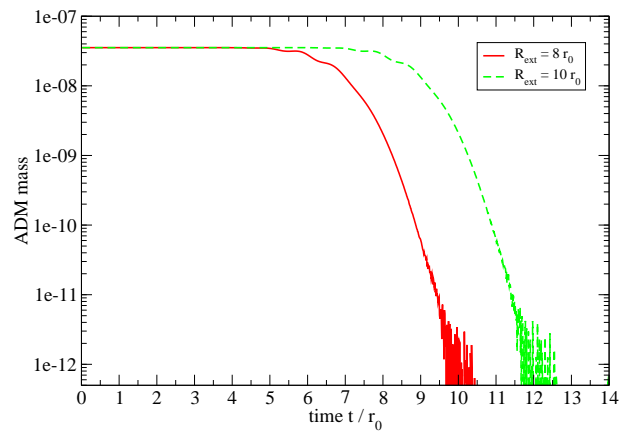} 
\caption{ \label{f:ADM2}
Evolution of the ADM mass (logarithmic scale, contrary to Fig.~\ref{f:ADM1})
for two different values of the wave extraction radius $R_{\rm ext}$.}
\end{figure}

\subsection{Results}

The time evolution of the component $h^{\ph\ph}$ of $\w{h}$ is
shown in Fig.~\ref{f:hpp}. All the
wave packet leaves the computational domain $r<8r_0$ around $t\sim 8r_0$
and we do not notice on Fig.~\ref{f:hpp} any spurious reflexion.

In order to test the code, we have monitored the ADM mass defined by
\be
    M_{\rm ADM} = {1\over16\pi} \oint_\infty 
    \left[ \cD^j \gm_{ij} - \cD_i \left( f^{kl} \gm_{kl} \right) 
    \right] dS^i , 
\ee
where the integral is taken over a sphere of radius $r=+\infty$ and
where we have adapted the original definition 
\cite{ArnowDM62} to general coordinates (i.e. non asymptotically Cartesian)
by the explicit introduction of the flat metric $\w{f}$. 
The above integral can be re-written in terms of 
the conformal metric and conformal factor:
\be
    M_{\rm ADM} = - {1\over16\pi} \oint_\infty \left(
        8 \cD_i \Psi + f_{ij} \cD_k h^{jk} - \cD_i h \right) dS^i .
\ee
Within Dirac gauge, the second term in the integrand vanishes identically,
whereas the last one does not contribute to the integral, due to the fast
decay of $h$ (at least $O(r^{-2})$) implied by Eq.~(\ref{e:det_unit_ortho1}).
Therefore the expression for the ADM mass reduces to the flux of the
gradient of the conformal factor:
\be
    M_{\rm ADM} = - {1\over2\pi} \oint_\infty 
        \cD_i \Psi\,  dS^i . \label{e:M_ADM}
\ee
Hence the expression of ADM mass in Dirac gauge is identical to the well
known expression for conformally flat hypersurfaces. 
The evolution of the ADM mass computed by means of Eq.~(\ref{e:M_ADM})
(let us recall that the sphere at $r=\infty$ belongs to our computational
domain) is presented in Fig.~\ref{f:ADM1}. For $t< 3r_0$, one sees that 
the ADM mass is conserved, as it should be, with an accuracy of four digits. 
Moreover, 
Fig.~\ref{f:ADM1} shows that the main source of error in the ADM mass
is the finite value of the time step $\delta t$.
For $t>3r_0$, the ADM mass starts to decrease, reflecting the fact 
that that the wave is leaving the domain $r\leq R_{\rm ext}$. Note that
by increasing the wave extraction radius from $R_{\rm ext}=8r_0$ to 
$R_{\rm ext}=10r_0$, we get a conservation of the ADM mass up to 
$t\simeq 5r_0$ (dashed curved in Fig.~\ref{f:ADM1}). 
In Fig.~\ref{f:ADM2}, we present the evolution 
of the ADM mass on a longer timescale. We see clearly that, after remaining
constant (the part shown in Fig.~\ref{f:ADM1}), the ADM mass decreases by
four orders of magnitude after $t\simeq 10r_0$ (resp. $t\simeq 12r_0$)
for the wave extraction radius $R_{\rm ext}=8r_0$ (resp. $R_{\rm ext}=10r_0$). 
The very small value of the ADM mass at late times demonstrates that all the wave packet
has leaved the domain $r\leq R_{\rm ext}$ 
and no spurious reflection has occurred.
This is due to the efficient outgoing wave boundary conditions 
\cite{NovakB04} set at the wave extraction radius.

\begin{figure}[htb]
\includegraphics[width=0.45\textwidth]{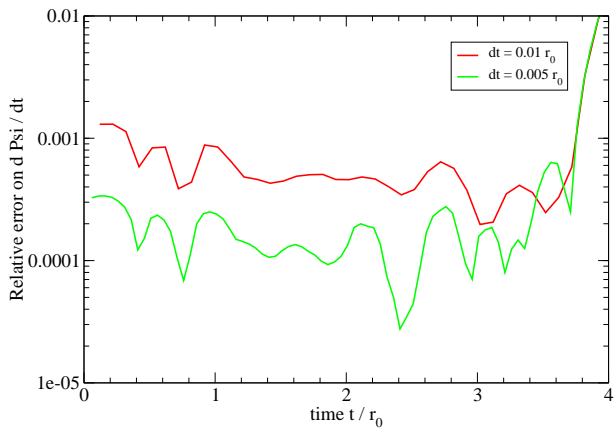} 
\caption{ \label{f:psidot}
Relative error $\varepsilon$ [Eq.~(\ref{e:error_psi_dot})]
on the time derivative of the conformal 
factor $\Psi$ in the central domain ($r\le r_0$).}
\end{figure}

\begin{figure}[htb]
\includegraphics[width=0.45\textwidth]{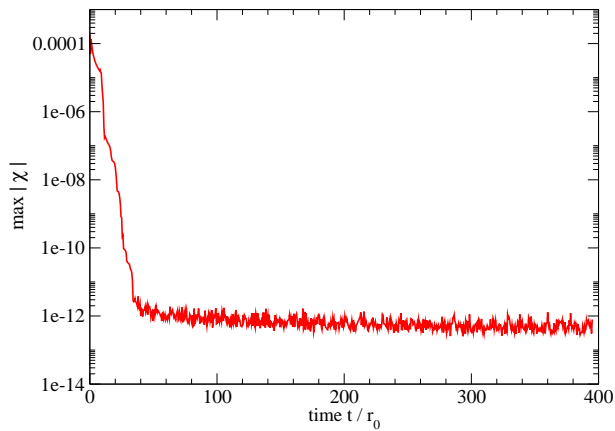} 
\caption{ \label{f:khimax}
Evolution of the maximum of absolute value of the potential $\chi$ 
[Eq.~(\ref{e:chi_h_def})] for the long term run.}
\end{figure}

Another test is provided by Eq.~(\ref{e:kin4_tr}) which relates the time 
derivative of the conformal factor $\Psi$ to the divergence of the shift 
vector $\w{\beta}$. As mentioned in Sec.~\ref{s:scheme}, this equation
must hold but is not enforced in our scheme. In a given numerical domain
we define the relative 
error on Eq.~(\ref{e:kin4_tr}) by
\be
\varepsilon := {
\left| \dert{\Phi}{t}  - \beta^k\cD_k\Phi -  {1\over 6} \cD_k\beta^k 
\right|
\over
{\rm max} \left| \dert{\Phi}{t} \right|
+ {\rm max} \left| \beta^k\cD_k\Phi + {1\over 6} \cD_k\beta^k 
\right| \label{e:error_psi_dot}
} ,
\ee
where the max are taken on the considered domain. We represent the value 
of $\varepsilon$ in the domain where it is the largest, namely the nucleus 
($r\le r_0$), in Fig.~\ref{f:psidot}. We see that 
Eq.~(\ref{e:kin4_tr}) is actually very well satisfied. The error is
in fact dominated by the time discretization (second order scheme), and
is as low as a few $10^{-4}$ for $\delta t = 5\; 10^{-3} r_0$. 
The increase
of $\varepsilon$ at $t\sim 4r_0$ is spurious and is due to the arrival
of the wave packet in the wave extraction domain $6r_0\le r\le 8r_0$. 

To check the long term stability of the code, we have let it run 
well after the wave packet has leaved the area $r<8r_0$, namely 
until $t = 400 r_0$. This very long time scale is similar with 
that used in Ref.~\cite{BaumgS99} to assess the stability of the BSSN
scheme. We found no instability to develop. In particular the maximum
value of the potential $\chi$ remains at the round-off error value
of $10^{-12}$ that has been reached at $t\sim 40 r_0$ (see Fig.~\ref{f:khimax}).

\section{Summary and conclusions} \label{s:concl}

We have introduced on each hypersurface $t={\rm const}$ of the 3+1 formalism
a flat 3-metric $\w{f}$, in addition to the (physical) 3-metric 
$\w{\gm}$ induced by the spacetime 4-metric $\w{g}$, in such a way
that asymptotically both metrics coincide.
This allows us to define properly the 
conformal metric $\w{\tgm}$ and not to stick to Cartesian coordinates.
A flat metric is introduced anyway, more or less explicitly, when
performing numerical computations.
We have written the 3+1 equations entirely in terms of the covariant
derivative associated with the flat metric $\w{f}$.
 
The Dirac gauge is expressed simply in terms of this flat metric 
as the vanishing of the divergence {\em with respect to $\w{f}$}
of the conformal metric $\w{\tgm}$.
Moreover in spherical components, the Dirac gauge reduces the 
resolution of the equations for  $\w{\tgm}$ to two scalar wave
equations. The remaining four components $\tgm^{ij}$ are then obtained
from the condition $\det \tgm^{ij} = \det f^{ij}$ and the three components
of the Dirac condition $\cD_j \tgm^{ij} = 0$. 
This clearly shows that the gravitational field has two degrees of 
freedom and this exhibits the TT wave behavior of the metric at infinity. 
Let us stress that the usage of spherical coordinates and spherical
components is essential for the reduction to two scalar wave equations. 
To our knowledge, this is the first time that a {\em differential} gauge
is used to directly compute some of the metric components, thus decreasing the
number of PDE to be solved. 
Previously, this was done only for {\em algebraic} gauges (i.e. 
setting some of the metric components to zero). 

Contrary to e.g. the minimal distortion gauge \cite{SmarrY78a}
or the ``Gamma-driver'' gauge \cite{AlcubB01},  
the Dirac gauge completely fix the coordinates (up to some boundary
conditions) in the initial hypersurface $\Sigma_0$. 
This implies that initial data must be prepared within this gauge, 
which might be regarded as a drawback (for instance an analytic
expression for the Kerr solution is not known in Dirac gauge). 
On the contrary, an advantage of the full coordinate fixing is to allow to
compute stationary solutions by simply setting $\dert{}{t}=0$ in the
various equations. For instance, Shibata, Uryu and Friedman \cite{ShibaUF04}
have recently proposed to use the Dirac gauge to compute quasiequilibrium 
configurations of binary neutron stars. 

In addition to the Dirac gauge, the use of the maximal slicing
results in an elliptic equation for the lapse function. 
Another elliptic equation for the conformal factor $\Psi$ (or equivalently
for $Q:= \Psi^2 N$)
arises from the Hamiltonian constraint. 
The Dirac gauge itself, in conjunction with the momentum constraint, 
results in an elliptic equation for the
shift $\w{\beta}$. The maximal slicing relates the divergence of
$\w{\beta}$ to the time derivative of the conformal factor.

Solving the above equations implies that the four constraints are
fulfilled by the solution.  
As already mentioned in the Introduction, some authors have
proposed very recently a scheme in which the
constraints, re-written as time evolution equations, are satisfied
up to the time discretization errors \cite{GentlGKM03}. On the contrary, 
in our scheme the constraints are fulfilled within the
precision of the {\em space} discretization errors (which can be
very low with a modest computational cost, thanks to spectral methods). 

It is worth noticing that the five elliptic equations of the widely used 
Isenberg-Wilson-Mathews
approximation to General Relativity \cite{Isenb78,IsenbN80,WilsoM89} (see
also Ref.~\cite{FriedUS02}) are naturally recovered in our scheme
by simply setting $\w{h}=0$: they are the equations for $N$, $Q$
and $\w{\beta}$. 

We have demonstrated the viability of the proposed constrained scheme
by numerically computing the evolution of a gravitational wave
packet in a vacuum spacetime. The numerical evolution has been found
to be both very accurate and stable.  
We are also made confident by existing constrained schemes
for vector equations which have proved to be successful:
the divergence-free hydro scheme of Ref.~\cite{VillaB02} (the constraint
being that the velocity field is divergence-free) and 
some MHD schemes in cylindrical coordinates \cite{KeppeT99} (the constraint
being that the magnetic field is divergence-free).

In this paper we have focused on space slices with $\mathbb{R}^3$
topology, except for Appendix~\ref{s:excision} where we briefly
discuss the properties of degenerate second order
operators and the number of boundary conditions at the surface
of excised holes with vanishing lapse. In a future work, we shall
focus on black hole spacetimes.

\begin{acknowledgments}
We warmly thank Thomas Baumgarte, Brandon Carter, Thibault Damour, 
John Friedman, Rony Keppens, Vincent Moncrief, Takashi Nakamura
and Koji Uryu for very fruitful discussions. We also thank the 
anonymous referee for pointing out some bibliographic references.
\end{acknowledgments}

\appendix

\section{Degenerate elliptic operators on a black hole horizon}
\label{s:excision}

In our view, a numerical scheme for black hole spacetimes should
recover known stationary solutions in coordinate-time independent
form (i.e. with the $\partial /\partial t$ coordinate vector
coinciding with the Killing vector of stationarity). 
Indeed we require arbitrary long term evolution of steady  state,
or quasi-steady state, black hole spacetimes.
For classical solutions (Kerr) in usual coordinates, this
requirement results in a vanishing lapse on the horizon
(see discussion in Refs.~\cite{GourgGB02,HannaECB03}). 
Therefore we excise from our computational domain a sphere $\cal H$
(or two spheres for binary systems) with $N=0$ as a boundary condition 
on that sphere and choose spherical coordinates such that $r=1$
on $\cal H$ \footnote{For a binary system, we introduce two 
coordinate systems, each centered on one hole, cf. \cite{GrandGB02}}.

In this case, the spatial operator acting on $\w{h}$ in 
Eq.~(\ref{e:wave_hij}) must not be merely the Laplacian $\Delta$
but 
\be
   \blacktriangle h^{ij} := N \Delta h^{ij} - \cD_k N 
   \left( \cD^i h^{jk} + \cD^j h^{ik} 
	 - \cD^k h^{ij}  \right) . 
\ee
This operator is formed by writing $\cD_k Q = \Psi^2\cD_k N + 2N\Psi\cD_k \Psi$
in the right-hand side of Eq.~(\ref{e:wave_hij}) and gathering the 
$\cD_k N$ term with the $\Delta h^{ij}$ one. 
The operator $\blacktriangle$ is degenerate,
because of the vanishing of $N$ at the boundary $\cal H$.
Similarly, the operator
acting on the shift vector $\w{\beta}$ is degenerate on $\cal H$
(cf. Eq.~(\ref{e:Poisson_beta}) with $A^{ij}$ given by 
Eq.~(\ref{e:Aij_calcul})
which contains a division by the lapse $N$).
Letting the unknown $u$ be a component of $h^{ij}$ or $\beta^i$,
these equations are of the kind
\be \label{e:poisson_degenere}
	N \Delta u + \epsilon \, \cD_i N \cD^i u = S , 
\ee
with the associated homogeneous equation
\be \label{e:degenere_hom}
	N \Delta u + \epsilon \, \cD_i N \cD^i u = 0 , 
\ee
where $N=0$ and $\partial N/\partial r > 0$ 
at $r=1$,
$\epsilon=\pm 1$ and $S$ is some effective source.
Since Eq.~(\ref{e:poisson_degenere}) is linear, a solution is 
a linear combination of a particular solution and a
homogeneous solution, i.e. a solution of Eq.~(\ref{e:degenere_hom}).
In the non-degenerate case, since Eq.~(\ref{e:degenere_hom}) is of second 
order, we have
two independent homogeneous solutions, which allow us to impose two 
boundary conditions. 
In the degenerate case ($N=0$ at $r=1$), the number of regular 
homogeneous solutions
depends upon the sign of $\epsilon$: two for $\epsilon = -1$ and only 
one for $\epsilon = +1$. To illustrate this, let us consider the following 
one-dimensional second order equation analogue to Eq.~(\ref{e:degenere_hom})
with $x=r-1$:
\be
	x {d^2 u\over dx^2}+ \epsilon {d u\over dx} = 0 ,\qquad \mbox{with}
	\ x\in [0,1] . 
\ee
The involved second-order operator is clearly degenerate at $x=0$. 
For $\epsilon=-1$, we have two independent homogeneous solutions:
\be
	u_1(x) = {\rm const} \qquad \mbox{and} \qquad u_2(x) = x^2 ,
\ee
whereas for $\epsilon=1$, the two independent homogeneous solutions are
\be
	u_1(x) = {\rm const} \qquad \mbox{and} \qquad 
	u_2(x) = \ln x . 
\ee
The last one is clearly not regular at $x=0$, so that in this case,
one can use only one homogeneous solution to satisfy a Dirichlet
boundary condition. 

This behavior of the degenerate operator can also be understood by
considering the parabolic (heat-like) equation 
associated with Eq.~(\ref{e:degenere_hom}):
\be \label{e:degenere_heat}
   {\partial u\over\partial t} = N \Delta u + \epsilon \, \cD_i N \cD^i u . 
\ee
The solution of the elliptic equation (\ref{e:degenere_hom}) is the
eigenfunction corresponding to the zero eigenvalue  of the spatial operator
acting on the right-hand side of Eq.~(\ref{e:degenere_heat}). In other words,
the solution $u$ we search for is the relaxed solution of the heat-like
equation (\ref{e:degenere_heat}). When $N\rightarrow 0$, 
Eq.~(\ref{e:degenere_heat}) becomes an advection equation near $r=1$, 
for which the number of boundary conditions at $r=1$ is zero or one
depending whether the ``effective velocity'' $-\epsilon \, \cD_i N = -
\epsilon \, {\partial N \over \partial r} \w{e}_r$ is ingoing or
outgoing at the boundary $r=1$.

For the spherical components of the shift vector, we have $\epsilon=-1$,
so that a boundary condition can always be given at $r=1$, in
addition to the boundary  condition at $r=\infty$.  
Regarding the spherical components of the metric potential $h^{ij}$, 
$\epsilon=1$ for $h^{rr}$, which means that no boundary condition
can be set at $r=1$ in addition to $h^{rr}=0$ at $r=\infty$.
On the contrary, $\epsilon = -1$ for the potential $\mu$ introduced
in Eqs.~(\ref{e:def_eta_mu_t})-(\ref{e:def_eta_mu_p}).
These points shall be studied more in details in a future work.
It is worth to mention that the boundary conditions for $h^{ij}$ at $r=1$
determine fully the coordinates within the Dirac gauge.

\end{document}